\documentclass[lettersize,journal]{IEEEtran}

\usepackage{amsmath,amssymb,amsfonts}
\usepackage{algorithmic}
\usepackage{algorithm}
\usepackage{array}
\usepackage{textcomp}
\usepackage{stfloats}
\usepackage{subfigure}    
\usepackage{graphicx}
\usepackage{xcolor}
\usepackage{bm}
\usepackage{cases}
\usepackage{xcolor}
\usepackage{url}
\usepackage{verbatim}
\usepackage{graphicx}
\usepackage{cite}
\newtheorem{lemma}{Lemma}[section]
\newtheorem{theorem}{Theorem}[section]
\begin{document}

\title{Secure Beamforming Design for IRS-ISAC Systems with a Hardware-Efficient Hybrid Beamforming Architecture}

\author{Weijie Xiong, Zhenglan Zhao, Jingran Lin, Zhiling Xiao, and Qiang Li
        \thanks{Copyright (c) 20xx IEEE. Personal use of this material is permitted. However, permission to use this material for any other purposes must be obtained from the IEEE by sending a request to pubs-permissions@ieee.org.}
\thanks{This work was supported in part by the Natural Science Foundation of China (NSFC) under Grant 62171110. \textit{(Corresponding author: Jingran Lin.)}.}
\thanks{Jingran Lin and Qiang Li are with the School of Information and Communication Engineering, University of Electronic Science and Technology of China, Chengdu 611731, China, the Laboratory of Electromagnetic Space Cognition and Intelligent Control, Beijing 100083, China, and the Tianfu Jiangxi Laboratory, Chengdu, Sichuan 641419, China (e-mail: jingranlin@uestc.edu.cn; lq@uestc.edu.cn).}
\thanks{Weijie Xiong, Zhenglan Zhao, and Zhiling Xiao are with the School of Information and Communication Engineering, University of Electronic Science and Technology of China, Chengdu 611731, China (e-mail: 202311012313@std.uestc.edu.cn; 202322010816@std.uestc.edu.cn; xiaozhiling@std.uestc.edu.cn).}}

\markboth{Journal of \LaTeX\ Class Files,~Vol.~14, No.~8, August~2021}%
{Shell \MakeLowercase{\textit{et al.}}: A Sample Article Using IEEEtran.cls for IEEE Journals}


\maketitle

\begin{abstract}
In this paper, we employ a hardware-efficient hybrid beamforming (HB) architecture to achieve balanced performance in an intelligent reflecting surface (IRS)-assisted integrated sensing and communication (ISAC) system. We consider a scenario where a multi-antenna, dual-function base station (BS) performs secure beamforming for a multi-antenna legitimate receiver while simultaneously detecting potential targets. Our objective is to maximize the communication secrecy gap by jointly optimizing the analog and digital beamformers, IRS reflection coefficients, and radar scaling factor, subject to constraints on beampattern similarity, total transmit power budget, and the constant modulus of both the analog beamformer and IRS reflection coefficients. This secrecy gap maximization problem is generally non-convex. To address this, we incorporate the exterior penalty method by adding the radar constraint as a penalty term in the objective function. We then propose an efficient approach based on the penalty dual decomposition (PDD) framework to solve the reformulated problem, featuring closed-form solutions at each step and guaranteeing convergence to a stationary point. Simulation results validate the effectiveness of the proposed algorithm and demonstrate the superiority of the IRS-ISAC system with HB architecture in balancing performance and hardware costs.
\end{abstract}


\section{Introduction}
Integrated sensing and communication (ISAC) systems not only allow communication and radar systems to share spectrum resources but also provide a fully integrated platform that transmits unified beamforming to simultaneously perform communication and radar sensing functions, significantly improving spectral, energy, and hardware efficiency \cite{liu2023seventy,zhang2021overview}. As a result, researchers from both academia and industry have explored various ISAC implementations \cite{liu2022joint,zhang2021enabling}.

Although the aforementioned designs significantly enhance the performance of secure beamforming for ISAC, these schemes may fail when the eavesdropping channel is correlated with the legitimate channel. In particular, if the communication user is close to the eavesdropper or if both are located in the same direction from the base station, the eavesdropping channel becomes highly correlated with the legitimate channel, leading to severe degradation in secure beamforming performance \cite{yang2023secure}. To tackle these issues, intelligent reflecting surfaces (IRS) have been developed and have gained considerable attention in recent years \cite{lu2020robust,wu2022robust}. An IRS is a software-controlled metasurface equipped with passive, digitally controlled reflecting elements that intelligently adjust the phase shift of impinging signals, creating a favorable propagation environment and providing additional optimization DoFs \cite{liu2023integrated,chepuri2023integrated,wang2021joint}. Specifically, by integrating IRS into ISAC systems, the correlation between the legitimate and eavesdropping channels can be proactively suppressed by dynamically adjusting the IRS phase shifts. Additionally, the reception quality for the communication user can be enhanced by utilizing the additional signals reflected by the IRS. Inspired by this, numerous studies have focused on designing secure beamforming for IRS-ISAC systems \cite{xing2023reconfigurable,kumar2023sca,hua2023secure,li2024noma}. For example, \cite{xing2023reconfigurable} initially investigated the impact of IRS on secure beamforming in a system with one legitimate user and an eavesdropper. This study ensured minimum communication performance for the legitimate user and demonstrated significant security improvements using IRS. However, the use of the interior point method (IPM) resulted in high computational complexity, making it unsuitable for real-time applications. To address this, \cite{kumar2023sca} proposed an algorithm based on the successive convex approximation (SCA) method, offering a closed-form solution for faster convergence. As a further step, \cite{hua2023secure} expanded the scenario to multiple legitimate users for the IRS-ISAC system, limiting maximum information leakage to the eavesdropper under both perfect and imperfect CSI conditions. To further enhance performance, \cite{li2024noma} introduced an AN-aided IRS-ISAC system using non-orthogonal multiple access (NOMA) technology. This approach improved secure beamforming performance by providing the system with more DoFs.

Despite the effectiveness of using IRS to combat eavesdropping in ISAC systems, the increased hardware costs and power consumption become significant, thus limiting their applicability. These increased costs arise from two main factors. Firstly, IRS-ISAC systems often employ fully-digital beamforming (FDB) architecture, which requires a dedicated radio frequency (RF) chain for each antenna. Secondly, an IRS typically consists of numerous reflecting elements made from materials with specific electromagnetic properties, such as metamaterials or phase-change materials. Therefore, it is essential to find a way to reduce hardware costs and power consumption while achieving balanced performance.

Recently, hybrid beamforming (HB) architecture has been successfully applied in ISAC systems to reduce hardware costs \cite{liu2019hybrid,kaushik2021hardware,liu2020joint}. This technology achieves near-optimal beamforming gain and directionality with low-cost phase shifters and fewer RF chains, providing an excellent trade-off between performance and complexity. For example, \cite{liu2019hybrid} was the first to evaluate the effectiveness of hybrid beamformers in reducing hardware costs for ISAC systems, demonstrating a favorable performance tradeoff between sensing and communication. Building on this, \cite{kaushik2021hardware} introduced a dynamic RF chain selection mechanism to optimize energy efficiency. Unlike the sub-connected HB architectures in \cite{liu2019hybrid,kaushik2021hardware}, fully-connected designs were employed in \cite{liu2020joint,cheng2021hybrid,qi2022hybrid}. For instance, \cite{liu2020joint} developed a low-complexity method for designing separate data streams for communication and sensing. In \cite{cheng2021hybrid}, a low-resolution hybrid beamforming with OFDM improved communication and radar accuracy. To enhance performance, \cite{qi2022hybrid} introduced a phase vector to provide additional DoFs for hybrid beamforming design and proposed an alternating minimization method.

Although HB architecture has been widely adopted in ISAC systems to reduce hardware costs and achieve various performance metrics, few studies have focused on its application in secure beamforming. Moreover, while the conventional HB architecture is well-suited for ISAC systems, its potential use in IRS-ISAC systems remains unexplored. Therefore, investigating how HB can enhance secure beamforming in IRS-ISAC systems is crucial, addressing both architectural and design mechanism perspectives. Inspired by the success of HB in ISAC systems, we aim to apply it to IRS-ISAC to balance secure beamforming performance and hardware complexity. This paper makes the following key contributions.
\begin{itemize}
\item Unlike previous studies on secure beamforming in IRS-ISAC systems that rely on fully digital implementations at the base station (BS), we propose a transmitter employing a fully-connected HB architecture. This approach reduces energy consumption and hardware costs by minimizing the need for a large number of RF chains.
\item We formulate the problem as a challenging non-convex optimization task aimed at maximizing communication secrecy gap. This is achieved by jointly optimizing the analog beamformer, digital beamformer, IRS reflection coefficients, and radar scaling factor, subject to constraints on beampattern similarity, total transmit power, and the constant modulus of both the analog beamformer and IRS reflection coefficients.
\item To efficiently solve this problem, we first incorporate the radar constraint into the objective function as a penalty term using the exterior penalty method. However, the reformulated problem remains non-convex and challenging to solve. We then propose an efficient algorithm based on the penalty dual decomposition (PDD) framework to solve the reformulated problem. The proposed PDD framework is highly efficient since it ensures closed-form solutions at each step while guaranteeing convergence to a stationary point.
\end{itemize}

The remainder of this paper is organized as follows: Section II introduces the system model and problem statement. In Section III, we reformulate the problem and propose a PDD-based algorithm to solve it. Simulation results are presented in Section IV, and Section V concludes the paper.

The following notations are used throughout the paper. A vector and a matrix are represented by $\bf a$ and $\bf A$ respectively; $(\cdot)^T$, $(\cdot)^H$ and $(\cdot)^*$ denote the transpose, conjugate transpose and conjugate respectively. $\bf I$ denotes an identity matrix with an appropriate dimension; ${\mathbb C}^N$ denotes the set of complex vectors of dimension $N$; the circularly symmetrix complex Gaussian distribution with mean $\mu $ and variance $\sigma^2 $ is denoted as $\mathcal{C N}(\mu, \sigma^2)$; $\text{Tr}({\bf A})$,  $||\cdot||_F$, $||\cdot||_2$ and  $|\cdot|$ represents trace operator, Euclidean norm, Frobenius norm and absolute value; $\text{diag}({\bf a})$ represnts a diagonal matrix where the elements of the vector $a$ are placed on the main diagonal; the phase of each element of a matrix is denoted as $\text{arg} ({\bf A})$; $\Re\{\cdot\}$ denotes the real part of a complex matrix; ${\bf A}\odot{\bf B}$ represents Kronecker product.

\section{System model and problem formulation}

\begin{figure}[!htbp]
  \begin{center}
  \includegraphics[width=3in]{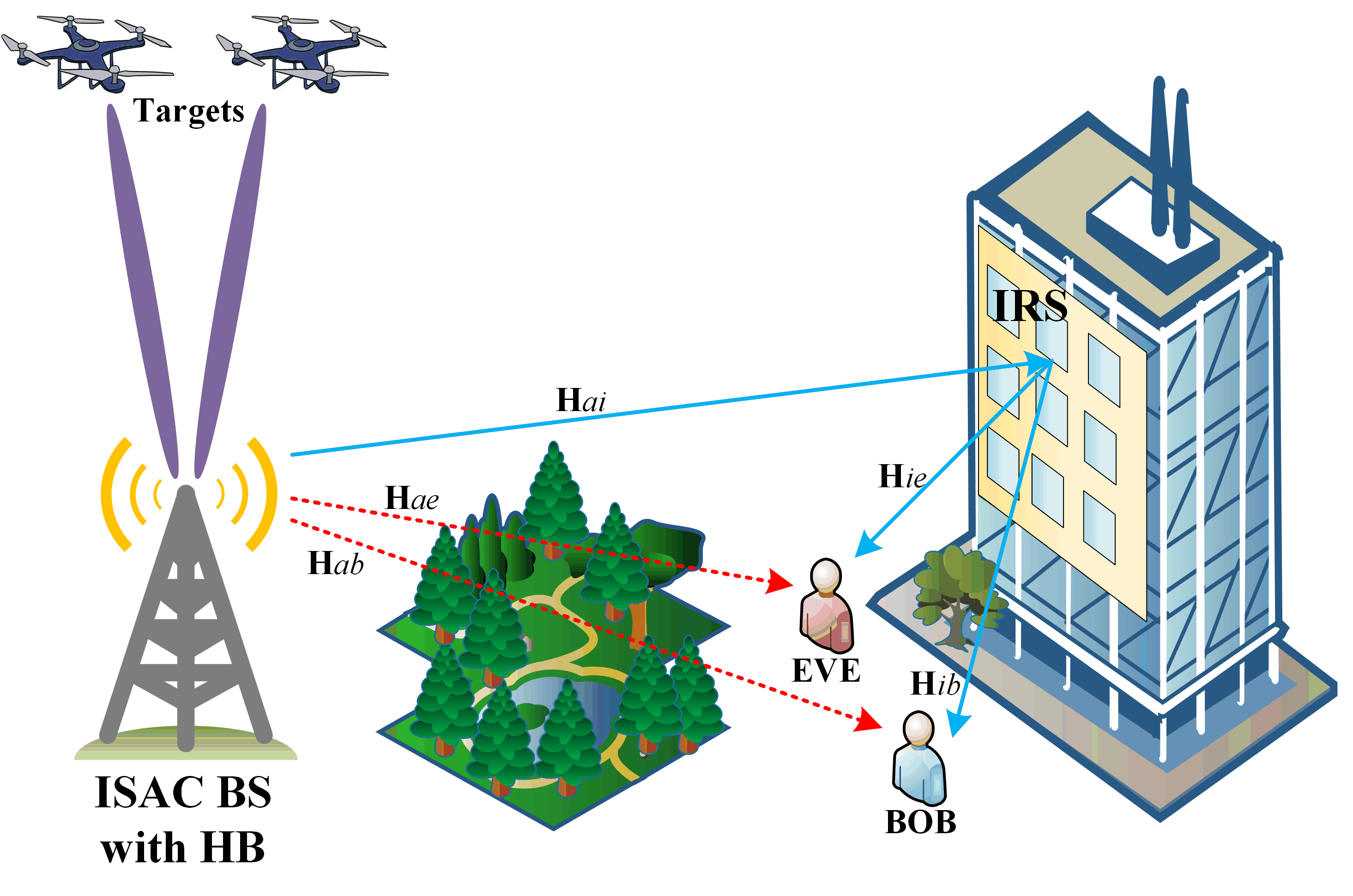}\\
  \caption{An IRS-ISAC system with HB for secure communication and radar target detection.}\label{IRS ISAC BS}
  \end{center}
\end{figure}

\begin{figure}[!htbp]
  \begin{center}
  \includegraphics[width=3in]{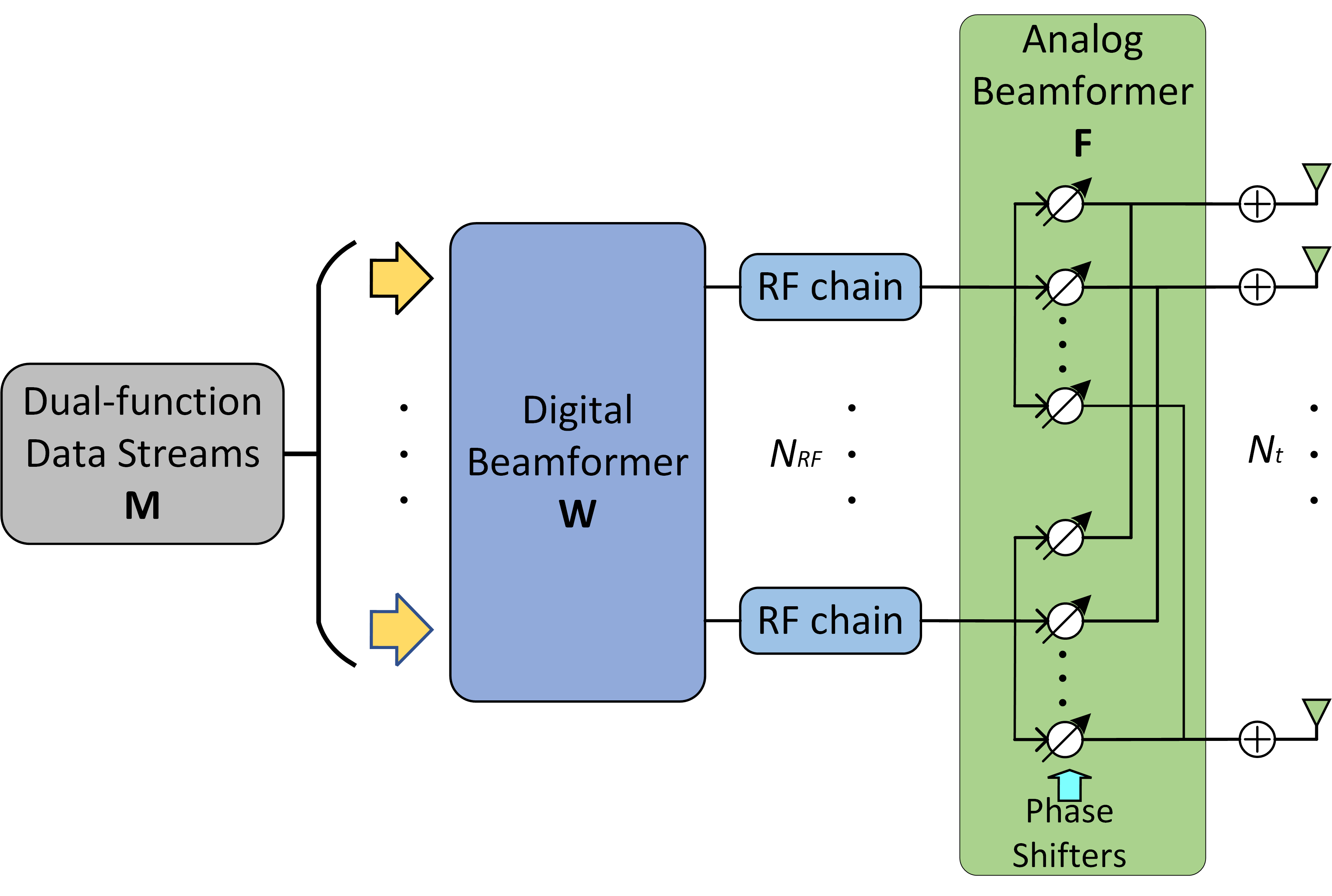} \\
  \caption{The architecture of fully-connected HB at the ISAC BS.}\label{HB}
  \end{center}
\end{figure}

As shown in Figure \ref{IRS ISAC BS}, we consider an ISAC BS employing a fully-connected HB architecture to serve an $N_b$-antenna legitimate communication user, referred to as Bob, assisted by an $N_i$-element IRS. In this scenario, an eavesdropper equipped with $N_e$ antennas, referred to as Eve, attempts to intercept the confidential data transmitted from the ISAC BS to Bob. Meanwhile, beamforming is also utilized to detect $T$ point-like targets located beyond the legitimate receiver. The IRS is positioned close to the communication user to effectively enhance downlink communication.

The fully-connected HB architecture is depicted in Figure \ref{HB}. Specifically, the digital baseband first generates data streams ${\bf s}(t) \in \mathbb{C}^{M}$ for both communication and radar sensing, with $M$ representing the lengths of data streams. Then, the data streams \textcolor{blue}{are} processed by an $N_{RF} \times M$ baseband digital beamformer ${\bf W} \in \mathbb{C}^{N_{RF} \times M}$, and then up-converted to the RF domain via $N_{RF}$ RF chains before being precoded with an $N_{t} \times N_{RF}$ analog beamformer ${\bf F} \in \mathbb{C}^{N_{t} \times N_{RF}}$. The analog  beamformer is implemented $N_{t}$ phase shifters, and thus subject to the magnitude constraint, i.e., $\left|{\bf F}\left[i,j\right]\right|=1,\forall i,j$. Then, the transmitted signal ${\bf x}(t) \in \mathbb{C}^{N_{t}}$ at the DFRC BS is given by \cite{liu2020joint},
\begin{equation}
    {\bf x}(t) =  {\bf F} {\bf W}{\bf s}(t). \label{inputsignal}
\end{equation} 

For the ISAC systems, we need a unified and tractable performance metric to describe the two tasks within one theme. To this end, inspired by \cite{luo2022joint} and \cite{yang2023secure}, we adopt the secrecy gap and the beampattern similarity as the performance measures of ISAC for secure beamforming and radar sensing, respectively. Let us first give a brief introduction of the two measures in the following.

\subsection{Secrecy Gap}
Assuming quasi-static channels, the received signals at Bob and Eve are respectively expressed as,
\begin{subequations}
\begin{align}
& {\bf y}_b(t)=({\bf H}_{ab}  +{\bf H}_{ib} {\bm \Phi  } {\bf H}_{ai}){\bf F} {\bf W} {\bf s}(t)+{\bf n}_b(t) \in \mathbb{C}^{N_{b}}, \\
& {\bf y}_e(t)= ({\bf H}_{ae} +{\bf H}_{ie} {\bm \Phi  }  {\bf H}_{ai}){\bf F} {\bf W} {\bf s}(t)+{\bf n}_e(t) \in \mathbb{C}^{N_{e}}, 
\end{align}
\label{receivesignal}
\end{subequations}
where $\mathbf{H}_{ab} \in \mathbb{C}^{N_b \times {N_{t}}}$,$\mathbf{H}_{ae} \in \mathbb{C}^{N_e \times {N_{t}}}$,$\mathbf{H}_{ai} \in \mathbb{C}^{{N_{i}} \times {N_t}}$,$\mathbf{H}_{ib} \in \mathbb{C}^{N_b \times N_i}$ and $\mathbf{H}_{ie} \in \mathbb{C}^{N_e \times N_i}$ are the channel matrices representing the direct link of BS-Bob, BS-Eve, BS-IRS, IRS-Bob and IRS-Eve respectively; ${\bf{n}}_b(t) \sim \mathcal{C N}(\mathbf{0}, \bf{I})$ and ${\bf{n}}_e(t) \sim \mathcal{C N}(\mathbf{0}, \bf{I})$ represent additive white Gaussian noise at Bob and Eve respectively; ${\bm \Phi  } = \text{diag}({\bm \phi}) \in \mathbb{C}^{N_i \times N_i} $ is the diagonal phase shift matrix for IRS; ${\bm \phi} = [\phi_1,\phi_2,...,\phi_{N_i}]^T \in \mathbb{C}^{N_i}$ denotes the reﬂection coefﬁcients.  

\textit{\textbf{Remark 1:}} In this paper, we leverage CSI differentiation to distinguish Bob and Eve \cite{wang2016physical}. Due to spatial separation and multipath effects, their CSI exhibits distinct characteristics, which can be analyzed through variations, statistical distributions, and temporal correlations. Additionally, RF fingerprinting, which exploits hardware-induced distortions, further enhances identification and ensures robust differentiation \cite{jagannath2022comprehensive}. Furthermore, we assume the transmitter has perfect CSI of Eve, which is feasible in certain scenarios. For instance, if Eve is also a user of the system, the transmitter may provide different services or content tailored to various user types, ensuring exclusivity for the target users. Additionally, for an active Eve, the CSI can be estimated from its transmissions. Interestingly, even for a passive Eve, it may be possible to estimate the CSI through the inadvertent leakage of local oscillator power from the RF front end of its receiver \cite{mukherjee2012detecting}.

The secrecy gap quantifies the difference in signal-to-noise ratio (SNR) between the legitimate communication channel and the eavesdropping channel. A higher SNR at the legitimate receiver, compared to the eavesdropper, implies stronger protection against unauthorized interception. This metric serves as an effective indicator of the system's ability to secure confidential information, as it reflects how well the legitimate user can decode the signal while minimizing the risk of eavesdropping. For the Gaussian wiretap channel, the secrecy gap can be calculated as \cite{li2019constant},
\begin{equation}
\begin{aligned}
{C_s}&=[ \text{SNR}_b  - \text{SNR}_e ]^{+}\\
&=\left[ ||{\bf H}_b({\bm \Phi }) {\bf F} {\bf W}||_F^2  - ||{\bf H}_e({\bm \Phi }) {\bf F} {\bf W}||_F^2 \right]^{+}, \label{SC}
\end{aligned}
\end{equation}
where $[\cdot]^{+} \triangleq \max (\cdot, 0)$; $\text{SNR}_b$ and $\text{SNR}_e$ denote the received SNR at the Bob and Eve, respectively; ${\bf H}_b({\bm \Phi  })={\bf H}_{ab} +{\bf H}_{ib} {\bm \Phi  } {\bf H}_{ai} \in \mathbb{C}^{N_b \times {N_{t}}}$ and ${\bf H}_e({\bm \Phi  })={\bf H}_{ae} +{\bf H}_{ie} {\bm \Phi  }  {\bf H}_{ai} \in \mathbb{C}^{N_e \times {N_{t}}}$.

\begin{figure}[!htbp]
  \begin{center}
  \includegraphics[width=3in]{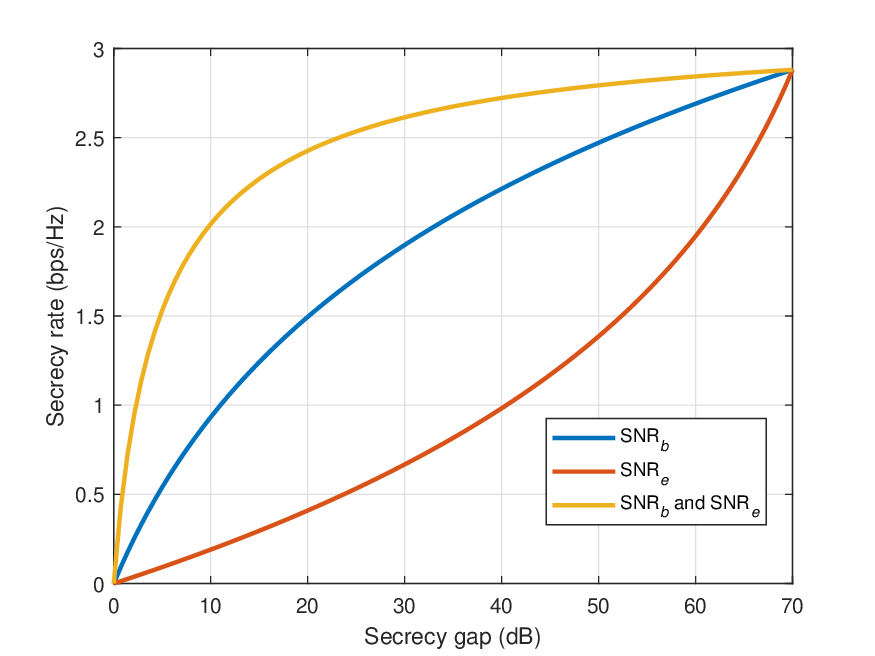}\\
  \caption{Secrecy rate versus secrecy gap for different SNR settings of Eve and Bob}\label{SNR}
  \end{center}
\end{figure}

\textit{\textbf{Remark 2:}} Most existing research uses the secrecy rate to evaluate secure performance, as given in \cite{ardestanizadeh2009wiretap},
\begin{equation}
{R_s} =[ \log_2(1+\text{SNR}_b)  - \log_2(1+\text{SNR}_e) ]^{+}
\end{equation}
Generally, an increase in secrecy rate indicates improved communication security. However, instead of using the secrecy rate, we use secrecy gap in this paper. Secrecy gap serves as an effective index for evaluating secure communication performance, as it provides a simpler yet meaningful measure of relative channel quality. A higher secrecy gap implies a stronger legitimate channel for Bob, correlating with a higher secrecy rate. Conversely, for Eve, a larger gap weakens the eavesdropping channel, making interception more difficult. Thus, secrecy gap effectively captures the system's ability to ensure secure data reception and is a practical proxy for assessing security. To further demonstrate the relationship between secrecy gap and secrecy rate, Figure \ref{SNR} illustrates the effect of varying SNR values. In the blue curve, Eve's SNR (\(\text{SNR}_e\)) is fixed while Bob's SNR (\(\text{SNR}_b\)) increases. In the red curve, Bob's SNR is fixed while Eve's SNR increases. In the yellow curve, both Bob's and Eve's SNR vary, with Bob's SNR increasing at a higher rate while Eve's SNR increases at a lower rate. The plot shows a positive relationship between secrecy gap and secrecy rate across all scenarios, confirming that a larger secrecy gap improves the secrecy rate and validating its use as a performance metric.

\subsection{Beampattern Similarity}
From the radar sensing perspective, we assume the IRS is deployed far away from the targets, which fly at low altitudes and have strong line-of-sight (LoS) links with the BS. Therefore, the radar transmit signals via the IRS are extremely weak and almost have no contribution to the detection of the target \cite{chu2023joint}. In order to pursue better target detection and estimation performance, a widely adopted approach is to maximize the signal power in the directions of the targets and minimize it elsewhere. In the sequel, the beampattern similarity metric that aims to match the designed beampattern with the ideal one is usually utilized to evaluate the sensing performance. Speciﬁcally, we deﬁne the steering vector for direction $\theta$ as,
\begin{equation}
{\bf{a}}( {\theta} ) = {[ {1,{{ e} ^{ j\pi \text{sin}{\theta}}},...,{{{e}} ^{ j\pi (N_t - 1)\text{sin}{\theta}}}} ]^T} \in {{\mathbb{C}}^{N_t \times 1}}.
\end{equation}
Then transmit beampattern can be expressed as \cite{li2007mimo},
\begin{equation}
\begin{aligned}
	P_b( \theta;{\bf F },{\bf W } ) &= \mathbb{E}\{| {\bf{a}}^H( {\theta} ) {\bf F} {\bf W}{\bf s}(t) |^2\} \\ &=  {\bf{a}}^H( {\theta} ){\bf F} {\bf W}{\bf W}^H {\bf F}^H{\bf{a}}( {\theta} ). 
\end{aligned}
\label{radarbeampattern}
\end{equation}
The mean squared error (MSE) between the ideal beampattern and the designed beampattern, which evaluates the beampattern similarity, is thus given by,
\begin{equation}
	P_r( {\delta,{\bf F },{\bf W } } )= \frac{1}{K}\sum\limits_{k = 1}^{K} {{| {\delta P_d(\theta_k) - P_b( \theta_k;{\bf F },{\bf W } )} |}^2},
 \label{MSE}
\end{equation}
where normalized $P_d(\theta_k)$ is the desired beampattern; $\theta_k$ denotes the $k$-th sampled angle; $\delta$ is a scaling factor. By introducing $\delta$, the designed beampattern approximates the appropriately scaled ideal beampattern, instead of $P_d(\theta_k)$ itself.

\textit{\textbf{Remark 3:}} While beampattern similarity is not a unique sensing metric, it remains a widely adopted and effective criterion in ISAC systems, as it directly evaluates the spatial power distribution without requiring explicit interference modeling in multi-target scenarios \cite{luo2022joint,chepuri2023integrated,liu2021dual}. In our considered system, inter-target interference does not need to be explicitly accounted for due to several factors. Firstly, high-resolution multi-antenna beamforming inherently limits energy leakage, ensuring that sensing beams remain highly directional and focused on intended targets. Secondly, the spatial separation between sensing targets and communication users further mitigates interference, as their operational regions do not significantly overlap. Thirdly, standard radar signal processing techniques, such as matched filtering, effectively suppress non-target signals, minimizing the impact of interference on target detection. Given these considerations, beampattern similarity remains a valid, interpretable, and tractable performance metric in multi-target systems.

\subsection{Problem Formulation}
Our goal is to maximize the communication secrecy gap by jointly optimizing the analog beamformer $\bf F$, the digital beamformer $\bf W$, the IRS reflection coefficients $\bm \phi$, and the radar scaling factor $\delta$, while satisfying constraints on beampattern similarity, total transmit power budget, and the constant modulus of both the analog beamformer and the IRS reflection coefficients. The optimization problem is thus formulated as,
 \begin{subequations}
\begin{align}
 \max _{{\delta},{\bf F}, {\bf W}, {\bm \Phi}} \quad &  C_s, \label{objective} \\ 
 \text { s.t. } \quad\quad & P_r( {\delta,{\bf F },{\bf W } } ) \le \varepsilon, \label{radarsimilarity} \\
&   ||{\bf F} {\bf W}||_F^2 \le P_{max}, \label{powerc} \\
&   \left|{\bf F}\left[i,j\right]\right|=1,\quad \forall i,j ,\label{CMab}\\
&   |{ \phi}_n|=1, \quad \forall n .\label{CMIRS}
\end{align}
\label{optimizationproblem}
\end{subequations}
where $\varepsilon$ represents the level of the beampattern similarity and $P_{max}$ denotes the available transmit power. The difficulty in optimizing problem (\ref{optimizationproblem}) can be attributed to several factors: firstly, the constraint (\ref{radarsimilarity}) involves a quartic term and is non-convex; secondly, the variables within the objective function (\ref{objective}) and the constraints (\ref{radarsimilarity}) and (\ref{powerc}) are highly coupled; and thirdly, the constraints (\ref{CMab}) and (\ref{CMIRS}) are non-convex due to constant modulus constraints. 

To address these challenges, in the following, we first incorporate the constraints (\ref{radarsimilarity}) into the objective function as a penalty term by utilizing the exterior penalty method. We then decouple the variables in (\ref{powerc}) by introducing an auxiliary variable. Finally, we propose an efficient method based on the penalty dual decomposition (PDD) framework to solve the problem.

\section{The proposed method}
\subsection{Problem Reformulation}
The objective function in problem (\ref{optimizationproblem}) evaluates the maximum between 0 and $\text{SNR}_b - \text{SNR}_e$, with only the latter term being variable-dependent. Therefore, after transforming the maximization into a minimization problem, we focus on the following equivalent formulation,
\begin{equation}
\begin{aligned}
 \min _{{\delta},{\bf F}, {\bf W}, {\bm \Phi}} \quad & ||{\bf H}_e({\bm \Phi }) {\bf F} {\bf W}||_F^2  - ||{\bf H}_b({\bm \Phi }) {\bf F} {\bf W}||_F^2, \\
 \text { s.t.} \quad\quad &(\ref{radarsimilarity}), (\ref{powerc}), (\ref{CMab}), (\ref{CMIRS}) \text { are satisfied }, \\
\end{aligned}
\label{minobjective}
\end{equation}

To simplify solving the non-convex quartic constraint in (\ref{radarsimilarity}), we employ the exterior penalty method \cite{CAO2022108644} by incorporating it as a penalty term into the objective function. This allows for a more flexible handling of the quartic constraint (\ref{radarsimilarity}) while maintaining the remaining problem as a constrained optimization, given by,
\begin{equation}
\begin{aligned}
 \min _{ {\delta}, {\bf F}, {\bf W}, {\bm \Phi}}  & \left\{\begin{array}{l} \mu(||{\bf H}_e({\bm \Phi }) {\bf F} {\bf W}||_F^2  - ||{\bf H}_b({\bm \Phi }) {\bf F} {\bf W}||_F^2)\\+(1-\mu)P_r( {{\bf F },{\bf W },\delta } ) \end{array}\right\}, \\
 \text {s.t.} \quad&(\ref{powerc}),  (\ref{CMab}), (\ref{CMIRS}) \text { are satisfied }, 
\end{aligned}
\label{reformulateproblem}
\end{equation}

Here, the penalty term is scaled by a parameter \(\mu\), which is iteratively increased to progressively enforce constraint satisfaction. Initially, \(\mu\) is set to a small value (e.g., \(\mu = 0.1\)), allowing flexibility in exploring the solution space. With each iteration, \(\mu\) is updated to \(\varsigma\mu\) (e.g., \(\varsigma = 1.1\)) to more heavily penalize constraint violations, ensuring they fall below a specified threshold (e.g., \(\varepsilon = 10^{-5}\)). This process strikes a balance between exploration and constraint adherence, guiding the solution toward feasibility while preserving the structure of other constraints. The exterior penalty method, detailed in Algorithm 1, was chosen for its simplicity and effectiveness in handling non-convex optimization constraints. Unlike interior-point methods, which require second-order derivatives and become computationally expensive for large-scale problems, this method reformulates constraints as penalty terms, enabling efficient and flexible solutions.

\begin{table}[!htbp] 
  \centering
  \begin{tabular}{rl}
   \hline
    \multicolumn{2}{l}{{\bf {Algorithm 1}}: Exterior penalty method to problem (\ref{minobjective})}  \\
    \hline
    0.& Initialize $0 \le \mu \le 1$, $\varsigma > 1$;\\
    1.& \textbf{\texttt{\textcolor{blue}{repeat}}} \\
    2.& \quad $\left\{{\delta},{\bf F}, {\bf W}, {\bm \Phi}\right\} \leftarrow \text{Optimize problem (\ref{reformulateproblem})}$ with updated $\mu$;\\ 
    3.& \quad  $\mu = \varsigma \cdot \mu$ \quad \textcolor{magenta}{\texttt{\%~update penalty parameter}}; \\
    4.& \textbf{\texttt{\textcolor{blue}{until}}} exterior penalty term $P_r( {\delta,{\bf F },{\bf W } } ) \le \varepsilon$ is satisfied. \\
    \hline
  \end{tabular}
\end{table}

Nevertheless, problem (\ref{reformulateproblem}) remains challenging due to the non-convexity introduced by the coupled variables in constraint (\ref{powerc}). To address this, we attempt to decouple $\bf F$ and $\bf W$ and reformulate the problem in a convex manner. To this end, we introduce an auxiliary variable ${\bf Q} \in \mathbb{C}^{N_t \times M} $ and recast problem (\ref{reformulateproblem}) as,
\begin{subequations}
\begin{align}
 \min _{{\delta}, {\bf F}, {\bf W}, {\bf Q}, {\bm \Phi}} \quad &\left\{\begin{array}{l} \mu(||{\bf H}_e({\bm \Phi }) {\bf Q}||_F^2-||{\bf H}_b({\bm \Phi }) {\bf Q}||_F^2)\\+  \frac{(1-\mu)}{K}\sum\limits_{k = 1}^{K} {\left|\begin{array}{l} \quad\quad\delta P_d(\theta_k) - \\{\bf a}^H( {\theta}_k ){\bf Q} {\bf Q}^H{\bf a}( {\theta}_k ) \end{array}\right|}^2  \end{array}\right\}, \label{objnewr} \\
 \text {s.t. } \quad\quad& ||{\bf Q}||_F^2  \le P_{max}, \label{powernewr}\\  
&  |{\bf F}[i,j]|=1,\quad \forall i,j ,\label{anlognewr}\\ 
&   |{\bm \phi}_n|=1, \quad \forall n,\label{IRSnewr}\\ 
&  {\bf Q} = {\bf F}{\bf W}.\label{equalc}
\end{align}
\label{reformulatedproblem2}
\end{subequations}
However, the reformulated problem remains non-convex and challenging to solve. With the highly coupled equality constraints in (\ref{equalc}) limiting the problem’s suitability for real-time distributed computation, the PDD method is a natural choice for addressing these challenges. It effectively decouples the constraints while ensuring the required consistency. In the following, we first provide a brief overview of the PDD method. Next, we demonstrate its application in handling the equality constraints in (\ref{equalc}). Finally, we design an efficient PDD-based algorithm to solve (\ref{reformulatedproblem2}).

\subsection{Brief Review of the PDD Method}
Essentially, the main idea of PDD is to dualize the complicated (or difﬁcult) coupling constraint by some proper penalty function, and then employ the coordinate descent computation to circularly update the variable blocks.

Consider the following problem,
\begin{equation}
\begin{aligned}
 \min _{{\bm \vartheta} , {\bm \iota}} \quad & \psi({\bm \vartheta}, {\bm \iota}) \triangleq {\bar \psi}({\bm \vartheta}, {\bm \iota})+\sum_{b=1}^B {\tilde \psi}({\bm \iota}_b), \\
 \text { s.t. } \quad &{\bm Y}({\bm \vartheta}, {\bm \iota})=0, \\
&  {\bm \Gamma}_p({\bm \vartheta}_p) \leq 0, {\bm \vartheta}_p \in {\bm \chi}_p ,p=1,2, \ldots, P,   \\
\end{aligned}
\label{decriptionproblem}
\end{equation}
where ${\bm \vartheta}=\{{\bm \vartheta}_p\}_{p=1}^P$ and ${\bm \iota}=\{{\bm \iota}_b\}_{b=1}^B$ are variables consisting of multiple blocks; ${\bar \psi}({\bm \vartheta}, {\bm \iota})$ is a scalar continuously differentiable
 function; ${\tilde \psi}({\bm \iota}_b)$ is a composite function taking the form of ${\hat \psi}_b({d}_b({\bm \iota}_b))$, with each ${d}_b({\bm \iota}_b)$ being a convex but possibly non-differentiable function, while ${\hat \psi}_b(\cdot)$ being a non-decreasing and continuously differentiable function; ${\bm Y}({\bm \vartheta}, {\bm \iota})$ and ${\bm \Gamma}_p({\bm \vartheta}_p)$ are vectors comprised of continuously functions. For more detailed description of problem (\ref{decriptionproblem}), the readers are referred to \cite{shi2020penalty}.

Besides the non-convexity and non-differentiability, the variable coupling with the equality constraint ${\bm Y}({\bm \vartheta}, {\bm \iota})$ further
complicates problem (\ref{decriptionproblem}). Without such a coupling constraint, block decomposition approaches can be utilized to efﬁciently decompose (\ref{decriptionproblem}) into multiple problems with lower dimension. This motivates the PDD method, which dualizes the constraint ${\bm Y}({\bm \vartheta}, {\bm \iota})$ as a proper penalty in the objective. For instance, following the augmented Lagrangian (AL) framework, we get the following problem \textit{(we reuse some notations in this subsection)},
\begin{equation}
\begin{aligned}
\left(\mathcal{P}_{{\bm \lambda},{\rho}} \right): \min _{{\bm \vartheta}, {\bm \iota}} \quad & {\cal L}_{\rho}({\bm \vartheta}, {\bm \iota},{\bm \lambda}) \triangleq  {\bar \psi}({\bm \vartheta}, {\bm \iota})+\sum_{b=1}^B {\tilde \psi}({\bm \iota}_b)\\&\quad\quad\quad\quad\quad+{\bm \lambda}^{H} {\bm Y}({\bm \vartheta}, {\bm \iota})+\frac{1}{2\rho}\|{\bm Y}({\bm \vartheta}, {\bm \iota})\|^2, \\
\text { s.t. }\quad & {\bm \Gamma}_p({\bm \vartheta}_p) \leq 0, {\bm \vartheta}_p \in {\bm \chi}_p, p=1,2, \ldots, P,
\end{aligned}
\end{equation}
where ${\bm \chi}$ is the Cartesian product of $P$ closed convex sets, i.e., ${\bm \chi} \triangleq \Pi_{p = 1}^P {{\bm \chi}_p}$; ${\bm \lambda}$ and $\rho > 0$ are the penalty parameter and dual variable with ${\bm Y}({\bm \vartheta}, {\bm \iota})=0$, respectively. Note that the resultant problem is completely separable among ${\bm \vartheta}_p$ and ${\bm \iota}_b$ for $p=1,2, \ldots,P$ and $b=1,2, \ldots,B$. Thereby, the problem can be efﬁciently solved in a coordinate descent manner.

The steps of PDD are summarized in Algorithm 2, that takes a double-loop structure, where the inner loop solves the AL subproblem $\left(\mathcal{P}_{{\bm \lambda}_{b_o},{\rho}_{b_o}} \right)$ approximately (i.e., Step 2), and the outer loop (with ${b_o}$ being the outer iteration index) updates the dual variable ${\bm \lambda}$ (i.e., Step 4), or the penalty parameter $\rho$ (i.e.,Step 8), according to the violation of constraint ${\bm Y}({\bm \vartheta}, {\bm \iota})=0$ (i.e., Step 3). Speciﬁcally, the notation ${\rm Optimize}\left(\mathcal{P}_{{\bm \lambda}_{b_o},{\rho}_{b_o}}; {\bm z}^{b_o-1}, \epsilon_{b_o} \right)$ in
Step 2 means calling some iterative algorithm to solve problem $\left(\mathcal{P}_{{\bm \lambda}_{b_o},{\rho}_{b_o}} \right)$, where ${\bm z}=({\bm \vartheta}, {\bm \iota})$ represents a feasible point of the problem. In Step 2, ${\bm z}^{b_o-1}$ is the initial point of iteration $b_o$, and $\epsilon_{b_o}$ denotes the accuracy of termination.
 \begin{table}[!htbp] 
  \centering
  \begin{tabular}{rl}
   \hline
    \multicolumn{2}{l}{{\bf {Algorithm 2}}: PDD-Based Solution to Problem (\ref{decriptionproblem})}  \\
    \hline
    0.& Initialize ${\bm z}^0$, $\rho_0 > 0$, ${\bm \lambda}_0$, $\kappa_0>0$, $\epsilon_0 > 0$, $\epsilon_{\rm stop} > 0$,\\
      & $0 < c < 1$ and $b_o=1$;\\
    1.& \textbf{\texttt{\textcolor{blue}{repeat}}} \\
    2.& \quad ${\bm z}^{b_o} = {\rm Optimize}\left(\mathcal{P}_{{\bm \lambda}_{b_o},{\rho}_{b_o}}; {\bm z}^{b_o-1}, \epsilon_{b_o} \right)$;\\ 
    3.& \quad \textbf{\texttt{\textcolor{blue}{if}}} $\|{\bm Y}({\bm z}^{b_o})\|_{\infty}  \leq \kappa_{b_o}$ \quad \textcolor{magenta}{\texttt{\%~update dual variables}}  \\
    4.& \qquad ${\bm \lambda}_{b_0+1} ={\bm \lambda}_{b_0} + \frac{1}{\rho_{b_o}} {\bm Y}({\bm z}^{b_o})$; \\
    5.& \qquad $ {\rho_{b_o+1}} = {\rho_{b_o}} $; \\
    6.& \quad \textbf{\texttt{\textcolor{blue}{else}}} \quad \textcolor{magenta}{\texttt{\%~update penalty parameter}} \\
    7.& \qquad ${\bm \lambda}_{b_0+1} ={\bm \lambda}_{b_0}$;\\
    8.& \qquad $ {\rho_{b_o+1}} = c \cdot {\rho_{b_o}} $; \\
    9.& \quad \textbf{\texttt{\textcolor{blue}{end}}} \texttt{\textcolor{blue}{if}} \\
   10.& \quad $\kappa_{b_o+1} = 0.9\times \|{\bm Y}({\bm z}^{b_o})\|_{\infty} $, and $\epsilon_{b_o+1} = 0.9\times\epsilon_{b_o}$; \\
   11.& \textbf{\texttt{\textcolor{blue}{until}}} some stopping criterion (e.g., $\|{\bm Y}({\bm z}^{b_o})\|_{\infty} \leq \epsilon_{\rm stop}$)\\
   &\qquad\qquad is satisfied. \\
    \hline
    
  \end{tabular}
\end{table}

\subsection{PDD-Based Algorithm for Problem (\ref{reformulatedproblem2})}
In this subsection, we fit problem (\ref{reformulatedproblem2}) into the PDD framework and design an efficient algorithm. To this end, we first introduce the Lagrange multiplier ${\bm \Psi} \in \mathbb{C}^{N_t \times M}$ and dualize the equality constraints (\ref{equalc}) using Lagrangian relaxation. This incorporates the constraints into the objective function, transforming them into penalty terms that balance objective minimization with constraint satisfaction. The AL function is constructed to unify objectives and constraints, guiding optimization toward feasible solutions. The AL function is given as,
\begin{equation}
\begin{aligned}
&{\cal L}_{\rho} ( {\delta}, {\bf F}, {\bf W}, {\bf Q}, {\bm \Phi}, {\bm \Psi} ) \triangleq \mu(||{\bf H}_e({\bf \Phi }) {\bf Q}||_F^2-||{\bf H}_b({\bf \Phi }) {\bf Q}||_F^2)\\&+\frac{(1-\mu)}{K}\sum\limits_{k = 1}^{K} {{| {\delta P_d(\theta_k) - {\bf a}^H( {\theta}_k ){\bf Q} {\bf Q}^H{\bf a}( {\theta}_k )} |}^2}  \\&  + \Re \{ \text{Tr}\{ {{\bm \Psi}}^H ( {\bf Q} - {\bf F}{\bf W} ) \} \}  + \frac{1}{2\rho} \|  {\bf Q} - {\bf F}{\bf W}  \|_F^2.
\end{aligned}
\end{equation}
Then, based on the PDD framework, in each outer iteration we ﬁrst solve the following problem to update $\{  {\delta}, {\bf F}, {\bf W}, {\bf Q}, {\bm \Phi} \}$, i.e., 
\begin{equation}
\begin{aligned}
 \min _{{\delta}, {\bf F}, {\bf W}, {\bf Q}, {\bm \Phi}} \quad &{\cal L}_{\rho} ( {\delta}, {\bf F}, {\bf W}, {\bf Q}, {\bm \Phi}, {\bm \Psi} ), \\
 \text {s.t.} \quad\quad\quad&(\ref{powernewr}), (\ref{anlognewr}), (\ref{IRSnewr}) \text{ are satisfied}, \\
\end{aligned}
\label{Lagrangian}
\end{equation}

Note that $\{  {\delta}, {\bf F}, {\bf W}, {\bf Q}, {\bm \Phi} \}$ are completely separable in the constraints of (\ref{Lagrangian}). Speciﬁcally, with the other variables ﬁxed, the resultant subproblem with respect to the remaining variable is relatively simple. Moreover, some subproblems can even be decomposed into multiple smaller-scale problems and solved efﬁciently. These observations motivates us to use the strategy of coordinate descent to iteratively solve (\ref{Lagrangian}), which constructs the inner loop of the algorithm. Since some subproblems in (\ref{Lagrangian}) may be non-convex, we adopt the block successive upper bound minimization (BSUM) method \cite{razaviyayn2013unified} to address this issue. Unlike the classic block coordinate descent (BCD) method, which directly minimizes the original objective, BSUM replaces the non-convex objective with a convex tight upper bound for each subproblem. This approximation ensures that the optimization remains tractable while retaining a close relationship to the original problem. Moreover, the BSUM framework guarantees stationary convergence under mild conditions, as established in Theorem 2 of \cite{razaviyayn2013unified}. These properties make BSUM particularly suitable for solving the challenging subproblems arising in our formulation. Next, we elaborate the details of updating ${\delta}$, ${\bf F}$, ${\bf W}$, ${\bf Q}$ and ${\bm \Phi}$.

\subsubsection{Update ${\delta}$} It can be easily found that the variable ${\delta}$ only exists in right-hand side of the objective function (\ref{Lagrangian}), of which is a quadratic and convex function with respect to the variable ${\delta}$. Therefore, by checking the first-order optimality condition \cite{boyd2004convex}, the optimal ${\delta}$ is expressed as,
\begin{equation}
\delta = \frac{\sum\limits_{k = 1}^{K} P_d(\theta_k) \text{vec}^H({\bf A}_k) \text{vec}({\bf Q}{\bf Q}^H) }{ \sum\limits_{k = 1}^{K} P_d(\theta_k)^2}, \label{soldelta}
\end{equation}
where ${\bf A}_k \buildrel \Delta \over= {\bf a}( {\theta}_k ){\bf a}^H( {\theta}_k ) \in \mathbb{C}^{N_t^2 \times N_t^2}$.

\subsubsection{Update ${\bf F}$} With the other variables being ﬁxed, the sub-problem related to $\bf F$ is given by,
\begin{subequations}
\begin{align}
\min _{ {\bf F}} \quad  & \Re \{ \text{Tr}\{ {\bm \Psi}^H ( {\bf Q} - {\bf F}{\bf W} ) \} \}  + \frac{1}{2\rho} \|  {\bf Q} - {\bf F}{\bf W}  \|_F^2,\label{objevtivesubF}\\
\text {s.t. } \quad& |{\bf F}[i,j]|=1,\quad \forall i,j. \label{constraintsubF}
\end{align}
\label{subF}
\end{subequations}
Due to the constant modulus constraints (\ref{objevtivesubF}), the sub-problem is still non-convex. To solve it efficiently, we try to find a tight upper bound of the objective function (\ref{objevtivesubF}) following the BSUM framework, which is given by, 
\begin{equation}
\begin{aligned}
\Re \{ \text{Tr}\{ {\bm \Psi}^H ( {\bf Q} &- {\bf F}{\bf W} ) \} \}  + \frac{1}{2\rho} \|  {\bf Q} - {\bf F}{\bf W}  \|_F^2 \\& \le \sum\limits_{i = 1}^{N_t} 2\Re \{ {\bf f}_i\left(({\bf G} -\lambda_{\text{max}}({\bf G}) {\bf I} ) ({\bf f}_i^k)^H - {\bf d}_i \right)\},
\end{aligned}
\label{maxsubF}
\end{equation}
where ${\bf f}_i^k \in \mathbb{C}^{1 \times N_{RF}}$ is the $i$-th row of the matrix at the $k$-th iteration $\bf F$; $\lambda_{\text{max}}({\bf G})$ denotes the maximum eigenvalue of $\bf G$; ${\bf G}={\bf W}{\bf W}^H \in \mathbb{C}^{N_{RF} \times N_{RF}}$; ${\bf D}={\bf W}({\bf Q} + \rho {\bf \Psi})^H= [{\bf d}_1,...,{\bf d}_i,...,{\bf d}_I] \in \mathbb{C}^{N_{RF} \times N_{t}}$. The detailed derivation is provided in Appendix \ref{appendixA}. 

Based on (\ref{maxsubF}), problem (\ref{subF}) can be further transformed into $N_t$ smaller problems related to ${\bf f}_i \in \mathbb{C}^{1 \times N_{RF}}$, $i=1,2,...,N_t$, respectively. The individual problem of ${\bf f}_i$ is expressed as,
\begin{subequations}
\begin{align}
 \min _{ {\bf f}_i} \quad &  2\Re \{ {\bf f}_i\left(({\bf G} -\lambda_{\text{max}}({\bf G}) {\bf I} ) ({\bf f}_i^k)^H - {\bf d}_i \right)\}, \label{tightupperboundF } \\
\text {s.t. }\quad & |{\bf f}_i(j)|=1,\quad \forall j.\label{CMCf}
\end{align}
\label{subreforF}
\end{subequations}
It is seen that problem (\ref{subreforF}) admits the following closed-form solution,
\begin{equation}
    {\bf f}_i = \text{exp}\left(j\text{arg}\left((\lambda_{\text{max}}({\bf G}) {\bf I}-{\bf G})({\bf f}_i^k)^H+{\bf d}_i\right)\right). \label{solveF}
\end{equation}

\begin{lemma} \label{lemma1}
(\ref{solveF}) is the unique optimal solution of the subproblem (\ref{subreforF}) although the constant modulus constraints related to $\bf f$ are generally non-convex.
\end{lemma}

\quad\textit{Proof:} Please refer to Appendix \ref{appendixD}.

\subsubsection{Update $\bf W$} With other variables being fixed, the subproblem related to $\bf W$ is an unconstrained convex problem, and can be expressed as,
\begin{equation}
    \min _{ {\bf W}} \quad \Re \{ \text{Tr}\{ {\bm \Psi}^H ( {\bf Q} - {\bf F}{\bf W} ) \} \}  + \frac{1}{2\rho} \|  {\bf Q} - {\bf F}{\bf W}  \|_F^2.
\end{equation}
By checking the first-order optimality condition, the optimal $\bf W$ is given as,
\begin{equation}
    {\bf W} = ({\bf F}^H{\bf F})^{-1}\left({\bf F}^H(\rho{\bm \Psi}+{\bf Q})\right). \label{solveW}
\end{equation}

\subsubsection{Update $\bf Q$} With other variables being fixed, the subproblem related to $\bf Q$ is given as,
\begin{subequations}
\begin{align}
 \min _{ {\bf Q}} \quad &\mu(||{\bf H}_e({\bf \Phi }) {\bf Q}||_F^2-||{\bf H}_b({\bf \Phi }) {\bf Q}||_F^2)\notag\\&+\frac{(1-\mu)}{K}\sum\limits_{k = 1}^{K} {{| {\delta P_d(\theta_k) - {\bf a}^H( {\theta}_k ){\bf Q} {\bf Q}^H{\bf a}( {\theta}_k )} |}^2} \notag\\ &+ \Re \{ \text{Tr}\{ {\bm \Psi}^H {\bf Q}   \} \}  + \frac{1}{2\rho} \|  {\bf Q} - {\bf F}{\bf W}  \|_F^2, \label{objecsubQ}\\
\text {s.t. } \quad& ||{\bf Q}||_F^2  \le P_{max},\label{conssubQ}
\end{align}
\label{subQ}
\end{subequations}
which remains non-convex due to the quartic and concave terms within the objective function. Similar to the update of $\bf F$, we follow the BSUM framework to find a tight upper bound of the objective function (\ref{objecsubQ}), and convert it into a convex one. After ignoring the constant terms, the majorized problem to solve problem (\ref{subQ}) for variable $\bf Q$ is formulated as,
\begin{subequations}
\begin{align}
 \min _{ {\bf Q}} \quad &\Re\{ \text{Tr}\{ {\bf Q}^H({\bf Z}_1{\bf Q}+ {\bf Z}_3 )  \} \} + \Re \{ \text{Tr}\{ {\bm \Psi}^H {\bf Q}   \} \}\notag \\ &+ \frac{1}{2\rho} \|  {\bf Q} - {\bf F}{\bf W}  \|_F^2, \label{objectsubQ}\\
\text {s.t. } \quad & ||{\bf Q}||_F^2  \le P_{max}.
\end{align}
\label{solQ}
\end{subequations}
where, 
\begin{equation}
\begin{aligned}
&{\bf Z}_1 = \mu {\bf H}_e^H({\bf \Phi }){\bf H}_e({\bf \Phi })+{\bf C}_1 \in \mathbb{C}^{N_t \times N_t},\\
&{\bf c}_1 = \text{vec}({\bf C}_1) = 2{\bf C}\text{vec}({\bf Q}^k({\bf Q}^k)^H) \in \mathbb{C}^{N_t^2},\\
&{\bf Z}_2 = -\mu{\bf H}_b^H({\bf \Phi }){\bf H}_b({\bf \Phi })-{\bf C}_2 - {\bf B}_t \in \mathbb{C}^{N_t \times N_t} ,\\
&{\bf c}_2 = \text{vec}({\bf C}_2) = 2\lambda_{\text{max}}({\bf C})\text{vec}({\bf Q}^k({\bf Q}^k)^H) \in \mathbb{C}^{N_t^2},\\
&{\bf b}_t  = \text{vec}({\bf B}_t) = \frac{(1-\mu)}{K}\sum\limits_{k = 1}^{K}2\delta P_d(\theta_k)\text{vec}({\bf A}_k) \in \mathbb{C}^{N_t^2},\\
 &{\bf C} = \frac{(1-\mu)}{K}\sum\limits_{k = 1}^{K}\text{vec}({\bf A}_k)\text{vec}({\bf A}_k)^H \in \mathbb{C}^{N_t^2 \times N_t^2},\\
&{\bf Z}_3 = 2 {\bf Z}_2^H {\bf Q}^k \in \mathbb{C}^{N_t \times M}.
\end{aligned}
\end{equation}
where ${\bf Q}^k $ is the matrix $\bf Q$ at $k$-th iteration; $\lambda_{\text{max}}({\bf C})$ denotes the maximum eigenvalue of $\bf C$. The detailed derivation is provided in Appendix \ref{appendixB}.

Now, the subproblem related to $\mathbf{Q}$ in (\ref{solQ}) becomes a convex problem. By checking the first-order optimality condition, the optimal $\mathbf{Q}$ is given by: 
\begin{equation}
{\bf Q} = \left(  2\rho{\bf Z}_1 + (2 \rho \alpha + 1) {\bf I}  \right)^{-1}({\bf F}{\bf W} - \rho {\bf Z}_2 - \rho {\bf \Psi}),
\end{equation}
where $\alpha$ is the Lagrangian multiplier chosen to satisfy the Karush-Kuhn-Tucker (KKT) conditions \cite{boyd2004convex}. To this end, we rewrite $2\rho{\bf Z}_1$ in the eigen-decomposition form, i.e,
\begin{equation}
2\rho{\bf Z}_1 = {\bf U} {\bm \Pi} {\bf U}^H,
\end{equation}
where ${\bm \Pi} = \text{Diag}\{{ \pi}_1,{ \pi}_2,...,{ \pi}_N  \} \in \mathbb{C}^{N_t \times N_t}$ with ${ \pi}_n $ being the $n$th nonnegative eigenvalue of $2\rho{\bf Z}_1$; ${\bf U}=[ {\bf u}_1,{\bf u}_2,...,{\bf u}_N ]\in \mathbb{C}^{N_t \times N_t}$ is the unitary matrix with ${\bf u}_n \in \mathbb{C}^{N_t}$ is the $n$th eigenvector of $2\rho{\bf Z}_1$. We further define,
\begin{equation}
{\bf \Delta} \buildrel \Delta \over = {\bf U}^H({\bf F}{\bf W} - \rho {\bf Z}_2 - \rho {\bf \Psi})({\bf F}{\bf W} - \rho {\bf Z}_2 - \rho {\bf \Psi})^H{\bf U},
\end{equation}
and have,
\begin{equation}
\|\mathbf{Q}\|_F^2 = \text{Tr}\{ \mathbf{Q} \mathbf{Q}^H \} = \sum_{n=1}^N \tfrac{\delta_n^n }{\left(\pi_n+(2\rho\alpha+1)\right)^2}, \label{ensurebisec}
\end{equation}
with $\delta_n^n$ being the $n$th diagonal element of ${\bf \Delta}$. Therefore, in the case of $\sum_{n=1}^N \tfrac{\delta_n^n }{\left(\pi_n+2\rho(1+\alpha)\right)^2} \le P$, we have $\alpha = 0$; otherwise, we find certain $\alpha > 0$ so that (\ref{ensurebisec}) holds for equality, which can be done by bisection, since $\|\mathbf{Q}\|_F^2$ is monotonically decreasing with $\alpha$.

To compute \(\mathbf{Q}\) by (\ref{ensurebisec}) via the bisection method, first evaluate \(S_0 = \sum_{n=1}^N \frac{\delta_n^n}{(\pi_n + 2\rho)^2}\). If \(S_0 \leq P_{max}\), set \(\alpha = 0\) as the power constraint is satisfied. Otherwise, initialize the bisection interval with \(\alpha_{\text{low}} = 0\) and choose an upper bound \(\alpha_{\text{high}}\) such that \(\sum_{n=1}^N \frac{\delta_n^n}{[\pi_n + 2\rho(1 + \alpha_{\text{high}})]^2} < P_{max}\). Iteratively compute the midpoint \(\alpha_{\text{mid}} = (\alpha_{\text{low}} + \alpha_{\text{high}})/2\) and evaluate \(S_{\text{mid}} = \sum_{n=1}^N \frac{\delta_n^n}{[\pi_n + 2\rho(1 + \alpha_{\text{mid}})]^2}\). If \(S_{\text{mid}} > P_{max}\), update \(\alpha_{\text{low}} = \alpha_{\text{mid}}\); otherwise, set \(\alpha_{\text{high}} = \alpha_{\text{mid}}\). Repeat this process until \(|S_{\text{mid}} - P_{max}|\) is within a predefined tolerance. Once the optimal \(\alpha^*\) is determined, substitute it back into equation (\ref{ensurebisec}) to obtain \(\mathbf{Q}\), thereby ensuring that \(\|\mathbf{Q}\|_F^2 = P_{max}\).

\subsubsection{Update $\bm \Phi$} Obviously, with other variables being fixed, the subproblem related to $\bm \Phi$ is given as,
\begin{subequations}
\begin{align}
 \min _{ {\bm \Phi}} \quad& ||{\bf H}_e({\bm \Phi }) {\bf Q}||_F^2-||{\bf H}_b({\bm \Phi }) {\bf Q}||_F^2, \\
\text {s.t. } \quad & |{\bm \phi}_n|=1, \quad \forall n.
\end{align}
\label{problemsPhi}
\end{subequations}
Since,
\begin{subequations}
\begin{align}
||{\bf H}_e({\bm \Phi }) {\bf Q}||_F^2 & = \text{Tr}\{{\bm \Phi}^H {\bf B}{\bm \Phi}{\bf E} + {\bm \Phi}^H{\bf J}^H + {\bm \Phi}{\bf J} \} \notag\\& = {\bm \phi}^H({\bf B} \odot {\bf E}^T){\bm \phi}+ 2 \Re\{ {\bm \phi}^H{\bf j}^*\},\\
||{\bf H}_b({\bm \Phi }) {\bf Q}||_F^2 & = \text{Tr}\{{\bm \Phi}^H {\bf M}{\bm \Phi}{\bf E} + {\bm \Phi}^H{\bf O}^H + {\bm \Phi}{\bf O} \} \notag\\& = {\bm \phi}^H({\bf M} \odot {\bf E}^T){\bm \phi}+ 2 \Re\{ {\bm \phi}^H{\bf o}^*\},
\end{align}
\label{expandforPhi}
\end{subequations}
where,
\begin{subequations}
\begin{align}
&{\bf B}= {\bf H}_{ie}^H{\bf H}_{ie} \in \mathbb{C}^{N_i \times N_i},\\
&{\bf E} = {\bf H}_{ai}{\bf Q}{\bf Q}^H {\bf H}_{ai}^H \in \mathbb{C}^{N_i \times N_i},\\
&{\bf j} = \text{diag}({\bf J})= {\bf H}_{ai}{\bf Q}{\bf Q}^H{\bf H}_{ae}^H{\bf H}_{ie} \in \mathbb{C}^{N_i },\\
&{\bf M}= {\bf H}_{ib}^H{\bf H}_{ib} \in \mathbb{C}^{N_i \times N_i},\\
&{\bf o} = \text{diag}({\bf O})= {\bf H}_{ai}{\bf Q}{\bf Q}^H{\bf H}_{ab}^H{\bf H}_{ib} \in \mathbb{C}^{N_i }.
\end{align}
\label{expandforPhimore}
\end{subequations}
Based on (\ref{expandforPhi}) and (\ref{expandforPhimore}), problem (\ref{problemsPhi}) can be reformulated as,
\begin{subequations}
\begin{align}
  \min _{ {\bm \phi}} \quad & {\bm \phi}^H\left(({\bf B}- {\bf M} )\odot {\bf E}^T\right){\bm \phi}+ 2 \Re\{ {\bm \phi}^H({\bf j}^*-{\bf o}^*)\}, \label{maxPhiobj}\\
\text {s.t. } \quad & |{\bm \phi}_n|=1, \quad \forall n,\label{maxPhicons}
\end{align}
\label{problem33new}%
\end{subequations}
which is still a non-convex problem due to the concave term within the objective function (\ref{maxPhiobj}) and the constant modulus constraints (\ref{maxPhicons}). Again, we try to find a tight upper bound of the objective function (\ref{maxPhiobj}) following the BSUM framework, which is given by, 
\begin{equation}
\begin{aligned}
{\bm \phi}^H&\left(({\bf B}- {\bf M} )\odot {\bf E}^T\right){\bm \phi}+ 2 \Re\{ {\bm \phi}^H({\bf j}^*-{\bf o}^*)\} \\& \le \Re \{ {\bf \phi}^H \left( ({\bf P} - \lambda_{max}({\bf P}){\bf I} ){\bf \phi}^k + ({\bf j}^*-{\bf o}^*) \right)  \},
\end{aligned}
\label{majorusenew}%
\end{equation}
where ${\bf P} = {\bf B}-{\bf M} \in \mathbb{C}^{N_i \times N_i}$; ${\bm \phi}^k \in \mathbb{C}^{N_i}$ is solution variable at the $k$-th iteration. The detailed derivation is provided in Appendix \ref{appendixC}. 

Based on (\ref{majorusenew}), the majorization problem of (\ref{problem33new}) is given as,
\begin{subequations}
\begin{align}
 \min _{ {\bf \phi}} \quad& \Re \{ {\bf \phi}^H \left( ({\bf P} - \lambda_{max}({\bf P}){\bf I} ){\bf \phi}^k + ({\bf j}^*-{\bf o}^*) \right)  \},\\
\text {s.t. } \quad & |{\bm \phi}_n|=1, \quad \forall n,
\end{align}
\label{MMusephi}%
\end{subequations}
It is evident that the structure of problem (\ref{MMusephi}) is identical to that of problem (\ref{subreforF}). Therefore, based on Lemma \ref{lemma1}, the problem (\ref{MMusephi}) admits the following unique optimal solution,
\begin{equation}
    {\bf \phi} = \text{exp}\left(j\text{arg}\left((\lambda_{\text{max}}({\bf P}) {\bf I}-{\bf P})({\bf \phi}^k)^H+({\bf o}^*-{\bf j}^*)\right)\right).
    \label{solPhi}
\end{equation}

The complete PDD-based algorithm summarizing the steps is presented in Algorithm 3. In the following section, we analyze the computational complexity and convergence of the proposed PDD-based algorithm.

\textit{\textbf{Remark 4:}} Unlike traditional PDD methods, our algorithm is specifically tailored to the proposed problem and incorporates several key innovations. Firstly, we introduce a novel penalty formulation to effectively address the coupling between analog and digital beamformers. Secondly, a customized BSUM method is developed to update the analog beamformer, digital beamformer, and IRS reflection coefficients under CM constraints, leveraging a tight surrogate function. Thirdly, the algorithm integrates hybrid beamforming with IRS control, offering additional degrees of freedom to enhance system performance. Lastly, it provides closed-form solutions at each step with guaranteed convergence to a stationary point. These advancements significantly improve computational efficiency and solution quality, making the algorithm particularly well-suited for hybrid beamforming in IRS-ISAC systems and advancing the application of PDD in non-convex optimization for wireless communications.

\begin{table}[!htbp]
  \centering
  \begin{tabular}{rl}
   \hline
    \multicolumn{2}{l}{{\bf {Algorithm 3}}: PDD-Based Solution to Problem (12)}  \\
    \hline
    0.& Initialize ${\delta}$, ${\bf F}$, ${\bf W}$, ${\bf Q}$, ${\bm \Phi}$, ${\bm \Psi}$, $\rho > 0$, $\kappa>0$, $\epsilon > 0$, \\
      &$\epsilon_{\rm stop} > 0$, $0 < c < 1$ and ${\cal L}_{\rho}^{\text{new}} ({\delta}, {\bf F}, {\bf W}, {\bf Q}, {\bm \Phi}, {\bm \Psi} )$;\\
    1.& \textbf{\texttt{\textcolor{blue}{repeat}}} \quad \textcolor{magenta}{\texttt{\%~outer PDD loop}} \\
    2.&\quad \textbf{\texttt{\textcolor{blue}{repeat}}} \quad \textcolor{magenta}{\texttt{\%~inner BSUM loop -- solve (\ref{Lagrangian})}} \\
    3.& \quad\quad ${\cal L}_{\rho}^{\text{old}}(\cdot)={\cal L}_{\rho}^{\text{new}}(\cdot)$;\\
    4.& \quad\quad Update $\delta$ according to (\ref{soldelta});\\
    5.& \quad\quad Update ${\bf F}$ according to (\ref{solveF});\\
    6.& \quad\quad Update ${\bf W}$ according to (\ref{solveW});\\
    7.& \quad\quad Update ${\bf Q}$ according to (\ref{solQ});\\
    8.& \quad\quad Update ${\bf \phi}$ according to (\ref{solPhi});\\
    9.& \quad\quad Compute ${\cal L}_{\rho}^{\text{new}} ( {\delta}, {\bf F}, {\bf W}, {\bf Q}, {\bm \Phi}, {\bm \Psi} )$ according to (\ref{Lagrangian});\\
    10.&\quad \textbf{\texttt{\textcolor{blue}{until}}} $\tfrac{|{\cal L}_{\rho}^{\text{old}}(\cdot)-{\cal L}_{\rho}^{\text{new}}(\cdot)|}{|{\cal L}_{\rho}^{\text{old}}(\cdot)|}\le \epsilon$ \\
    11.&\quad $\text{error}= || {\bf Q}-{\bf F}{\bf W} ||_\infty$ \\;
    12.& \quad \textbf{\texttt{\textcolor{blue}{if}}} $\text{error}\le\kappa$ \quad \textcolor{magenta}{\texttt{\%~AL method -- Update ${\bm \Psi}$}}\\
    13.& \quad\quad ${\bm \Psi}={\bm \Psi}+\tfrac{1}{\rho}({\bf Q}-{\bf F}{\bf W})$;\\
    14.& \quad \textbf{\texttt{\textcolor{blue}{else}}} \quad \textcolor{magenta}{\texttt{\%~penalty method -- Update $\rho$}}\\
    15.& \quad\quad $\rho = c \rho$;\\
    16.& \quad \textbf{\texttt{\textcolor{blue}{end}}} \\
    17.& \quad $\kappa = 0.9 \times \text{error}$, $\epsilon = 0.9 \times \text{error}$ \\
    18.& \textbf{\texttt{\textcolor{blue}{until}}} $\text{error} \le \epsilon_{\rm stop}$ \\
   
    \hline
  \end{tabular}
\end{table}

\subsection{Analysis of Computation Complexity and Convergence}
\subsubsection{Analysis of computation complexity} The complexity of Algorithm 2 is primarily due to the inner BSUM loop, which is analyzed as follows. The update of $\delta$ requires about ${\cal O}({N_t^2 K})$ operations. The update of $\bf F$ requires about ${\cal O}({N_t N_{RF}^3})$ operations. The update of $\bf W$ requires about ${\cal O}({N_{RF}^3})$ operations. The update of $\bf Q$ requires about ${\cal O}({N_{t}^3})$ operations. The update of $\bm \phi$ requires about ${\cal O}({N_{i}^3})$ operations. Therefore, the total complexity of the PDD-based algorithm is approximately ${\cal O}\left( T_1 T_2 (N_t(N_t K + N_{RF}^3 + N_t^2)  +{N_{RF}^3}+{N_{i}^3})\right)$ operations, where $T_1$ and $T_2$ are the total number of BSUM iterations and PDD iterations, respectively.
 
\subsubsection{Analysis of convergence} To establish convergence, we first demonstrate that the inner loop BSUM method for solving the augmented Lagrangian (AL) problem (\ref{Lagrangian}) converges to a stationary point. Specifically, we rely on the following theorem from \cite{razaviyayn2013unified}.

\begin{theorem} \label{theorem1} Suppose that the surrogate function is quasi-convex and serves as a tight upper bound for the original subproblems with respect to ${\delta}$, ${\bf F}$, ${\bf W}$, ${\bf Q}$ and ${\bm \Phi}$. Furthermore, assume that each subproblem has a unique solution for any point within its constraints. Then, every limit point of the iterates generated by the BSUM algorithm is a coordinate-wise minimum. Additionally, if the original problem is regular, the BSUM algorithm converges to a stationary point. \end{theorem}

Since the inner loop BSUM method strictly follows Theorem \ref{theorem1}, its convergence is straightforward to establish. Furthermore, as the constraints (\ref{powernewr})-(\ref{equalc}) satisfy Robinson's condition \cite{bertsekas1997nonlinear}, it is stated in \cite{shi2020penalty} that every limit point generated by Algorithm 2 is a stationary point of the problem (\ref{reformulatedproblem2}).

\subsection{Discussion on Method Adaptability}
This subsection briefly evaluates the adaptability of the proposed method to sub-connected HB structures \cite{nguyen2019unequally,wan2021performance,gadiel2021dynamic}. While the proposed method is primarily designed for fully-connected HB, it demonstrates inherent flexibility, offering potential extensions to accommodate the sub-connected HB architectures.

The primary difference between the sub-connected and fully-connected HB architectures lies in the form of the analog beamformer \( \mathbf{F} \), which is represented as,
\begin{equation}
    [{\bf f}_1, {\bf f}_2, ..., {\bf f}_{N_{RF}}]\quad \text{and}\quad \begin{bmatrix}
\mathbf{f}_1 & 0 & \cdots & 0 \\
0 & \mathbf{f}_2 & \cdots & 0 \\
\vdots & \vdots & \ddots & \vdots \\
0 & 0 & \cdots & \mathbf{f}_{N_\text{RF}}
\end{bmatrix},
\end{equation}
for the fully-connected (left) and sub-connected (right) architectures, respectively. Here, \( {\bf f}_{j} \in \mathbb{C}^{\hat{N}_t}, j \in [N_{RF}] \), with \( \hat{N}_t = N_t \) for the fully-connected architecture and \( \hat{N}_t = \frac{N_t}{N_{RF}} \) for the sub-connected architecture. In both cases, \( \mathbf{F} \) has dimensions \( N_t \times N_{RF} \), but the fully-connected configuration employs a full matrix, whereas the sub-connected configuration adopts a block-diagonal matrix due to constrained connectivity. 

The sub-connected HB architecture can be viewed as a special case of the fully-connected HB architecture, where the off-diagonal elements of \( \mathbf{F} \) are zero. The proposed method can be adapted to the sub-connected HB architecture by incorporating a projection operation during the optimization of \( \mathbf{F} \) in (\ref{solveF}). This operation enforces the block-diagonal structure, ensuring compatibility with the sub-connected architecture.

\section{Numerical Results}
In this section, we present simulation results to evaluate the secure beamforming design for the proposed IRS-ISAC system with HB architecture. For comparison, the proposed method is compared to the following benchmark schemes: 
\begin{itemize}
    \item \textbf{AO-SDR-SAME}: the same IRS-ISAC system with the HB architecture is considered, but the problem is solved using the alternating optimization (AO) method \cite{xu2024enhancing}, which directly decouples the variables into several subproblems. Each subproblem is then solved using the semi-definite relaxation (SDR) method \cite{5447068};
    \item \textbf{IRS-C-HB}: the HB architecture with the assistant of the IRS only works for the downlink communication system \cite{yang2023secure};
    \item \textbf{WOIRS-C-FDB}: the HB architecture without the assistant of the IRS only works for the downlink communication system \cite{cheng2021hybrid};
    \item \textbf{WOIRS-C-FDB}: the FDB architecture without the assistant of the IRS only works for the downlink communication system \cite{li2019constant};
    \item \textbf{IRS-ISAC-FDB}: the FDB architecture for the ISAC system with the assistant of the IRS \cite{xing2023reconfigurable}; 
    \item \textbf{WOIRS-ISAC-FDB}: the FDB architecture for the ISAC system without the assistant of the IRS \cite{dong2023joint};
    \item \textbf{R-Optimal}: the MIMO radar system only works for radar sensing \cite{cheng2017constant}.
\end{itemize}

Unless otherwise specified, we assume that the ISAC BS is equipped with $N_t=10$ transmit antennas, $N_{RF}=4$ RF chains, and a data stream length of $M=2$. The transmit power budget is set as $P_{max} = 1$dB. The weight coefficient $\mu =0.5$. For communication, the number of the IRS elements is set as $N_i=32$, and the number of antennas at Eve and Bob to $N_b=4$ and $N_e=4$, respectively. Following \cite{dong2020enhancing}, we examine a fading environment where all channels comprise both large-scale and small-scale fading. Small-scale fading matrix entries are generated as complex zero-mean Gaussian random variables with unit covariance. The large-scale fading path loss is set at -30dB for a reference distance of 1m, with path loss exponents for all links at 3. We assume distances between Alice and Bob, Alice and IRS, Alice and Eve, IRS and Bob, and IRS and Eve to be 80m, 30m, 80m, 40m, and 40m, respectively. For radar sensing, the spatial area of interest is $\theta = [-90^{\circ},90^{\circ}]$ and we set the interval of $\{\theta_k\}_{k=1}^K$ as $1^{\circ}$ and $K = 181$. The desired beampattern $P_d(\theta_k)$ is given by,
\begin{equation}
	P_d(\theta_k) = \left\{ {\begin{array}{*{20}{c}}
			1, & \theta_k  \in [ {\bar \theta_t} - \tfrac{\Delta \theta}{2},{\bar \theta _t} + \tfrac{\Delta \theta}{2}],\\
			0, & {\text{otherwise}}.
	\end{array}} \right.
\end{equation}
where ${\bar \theta_t}$ represents the direction of the $t$-th target. We examine the desired beampattern with ${\bar \theta_1} = -40^{\circ}$, ${\bar \theta_2} = 0^{\circ}$, and ${\bar \theta_3} = 40^{\circ}$, each having a beam width of ${\Delta \theta} = 20^{\circ}$ unless otherwise specified. Additionally, the hyper-parameters are set as follows: $\varsigma=1.1$, $\mu=0.5$, $\rho = 0.1$, $\kappa = 0.9$, $\epsilon = 10^{-5}$, $\epsilon_{\rm stop} = 10^{-5}$ ,and $c = 0.7$. Simulation results are averaged over 100 random fading realizations.

\begin{figure}[t]
  \begin{center}
  \includegraphics[width=3in]{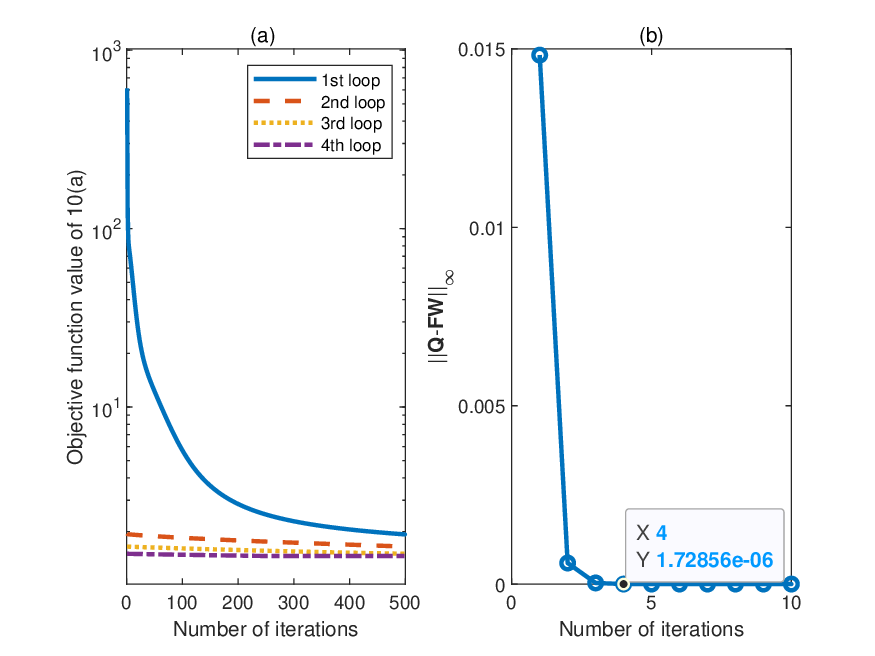}\\
  \caption{Convergence of the proposed PDD-based algorithm.}\label{converg}
  \end{center}
\end{figure}

\subsection{Analysis of convergence}

Figure \ref{converg} presents the convergence behavior of the proposed PDD-based algorithm. Figure \ref{converg}(a) depicts the objective function value against the number of iterations. It can be observed that the objective value (\ref{objnewr}) within each inner loop converges monotonically within a limited number of iterations, with the rate of decrease reducing sharply after the initial loop. Additionally, the objective value at the end of one loop is smaller than at the start of the next, indicating that the outer loop also converges monotonically. Figure \ref{converg}(b) shows the constraint violation $\|\mathbf{Q} - \mathbf{FW}\|_{\infty}$, demonstrating that the error of the equality constraint quickly converges within four iterations, confirming the feasibility of the solution obtained by the PDD-based algorithm for the original problem.

\begin{figure}[t]
  \begin{center}
  \includegraphics[width=3in]{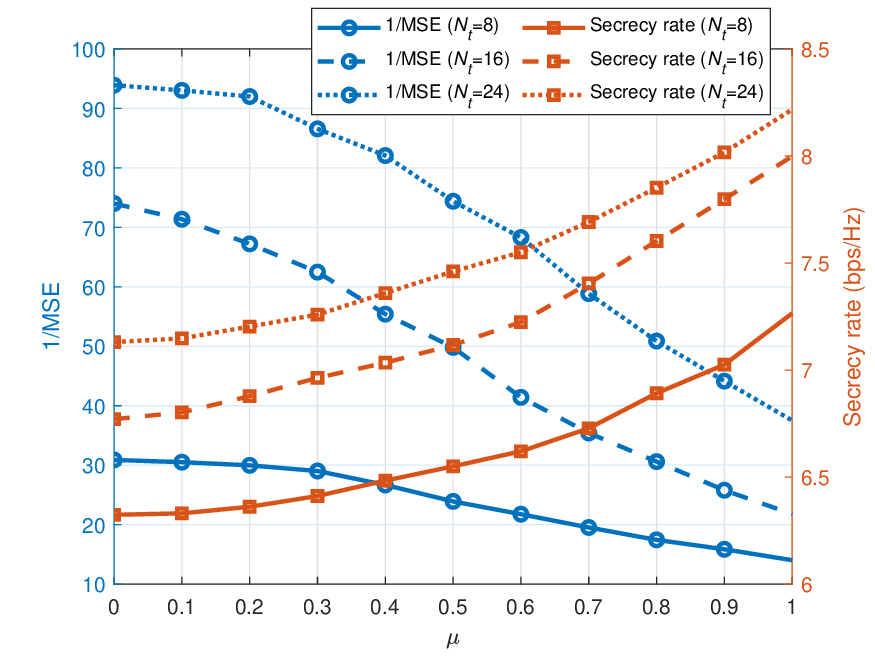}\\
  \caption{1/MSE and secrecy rate versus weight coefficient ($\mu$) as it increases from 0 to 1}\label{tradeoffp}
  \end{center}
\end{figure}

\subsection{Radar and Communication Trade-Off Analysis}

Figure \ref{tradeoffp} illustrates the trade-off between radar sensing performance (measured by the mean squared error (MSE) in (\ref{MSE})) and communication performance (measured by the secrecy rate) as the weight coefficient \(\mu\) varies from 0 to 1. To provide a clearer representation of the trade-off, \(1/\text{MSE}\) is plotted in the figure, as higher values of \(1/\text{MSE}\) indicate better radar sensing performance. The weight coefficient \(\mu\) serves as a critical parameter for balancing the two objectives. When \(\mu\) approaches 1, the system focuses more on communication, leading to a noticeable improvement in the secrecy rate. Conversely, as \(\mu\) approaches 0, the system prioritizes radar sensing, resulting in higher values of \(1/\text{MSE}\), which indicate enhanced radar sensing performance. This trade-off highlights the system’s ability to adapt to varying application requirements by adjusting the emphasis on either communication or radar sensing. Furthermore, as the number of transmit antennas (\(N_t\)) increases, both radar sensing and communication performance improve across all values of \(\mu\). This is due to the additional DoFs provided by the increased number of antennas.

\begin{figure}[t]
\centering
\subfigure[The beampattern for the first case.]{
\includegraphics[width=3in]{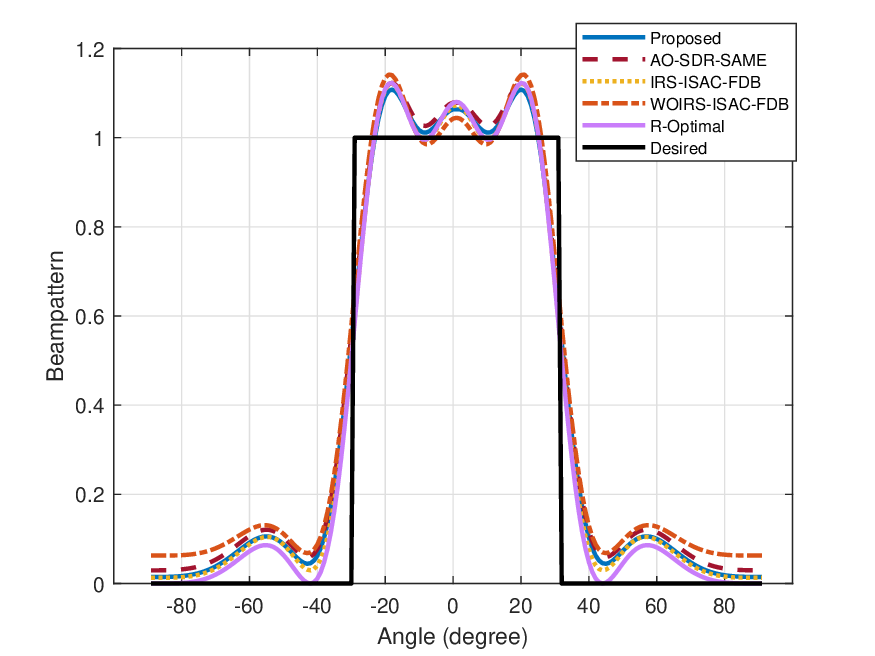} 
}
\subfigure[The beampattern for the second case.]{
\includegraphics[width=3in]{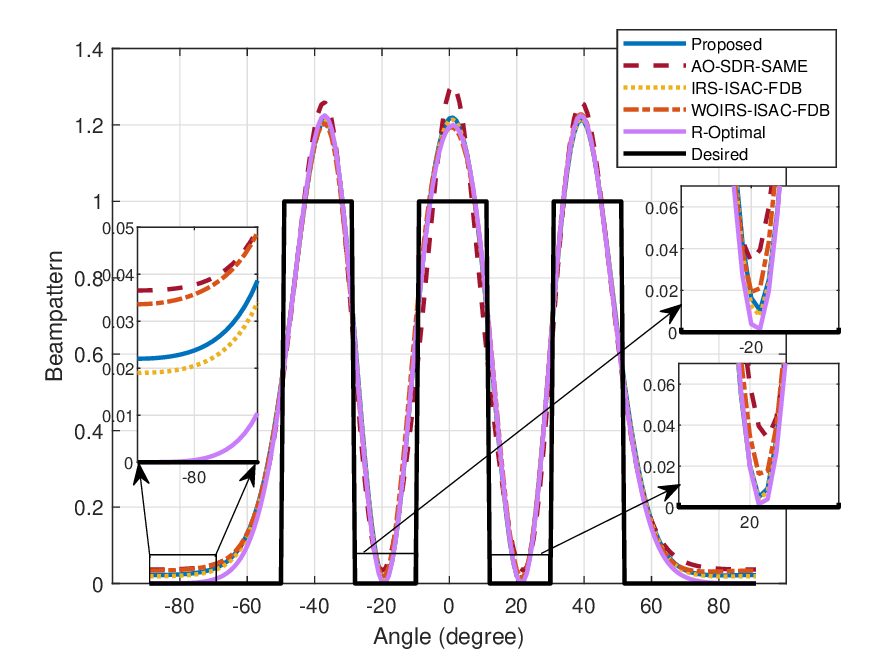} 
}
\DeclareGraphicsExtensions.
\caption{Comparison of beampatterns between different architectures and cases.}\label{beampgr}
\end{figure}

\subsection{Comparisons of Radar Performances}

In this subsection, we consider an additional desired beampattern with ${\bar \theta_1} = 0^{\circ}$ and a beam width of ${\Delta \theta} = 60^{\circ}$ as the first case, while the previously mentioned desired beampattern is treated as the second case to better illustrate the sensing performance.

Figure \ref{beampgr} compares the beampattern defined in (\ref{radarbeampattern}) across different architectures and scenarios. As shown in Figures \ref{beampgr}(a) and \ref{beampgr}(b), the proposed method achieves lower sidelobe levels compared to the 'WOIRS-ISAC-FDB' architecture, confirming that the ISAC system demonstrates improved sensing performance when utilizing IRS to assist communications. Additionally, the proposed method performs nearly as well as the 'IRS-ISAC-FDB' architecture, highlighting that the HB architecture is more hardware-efficient, as it requires only one-fourth of the RF chains compared to the FDB architecture. Furthermore, the proposed method exhibits lower sidelobe levels than the 'AO-SDR-SAME' architecture, confirming that the proposed PDD algorithm can converge to a better solution. Finally, the proposed method approaches the performance of the 'R-Optimal' architecture, with the gap between them being leveraged for secure communication.

\begin{figure}[t]
\centering
\subfigure[The MSE for the first case.]{
\includegraphics[width=3in]{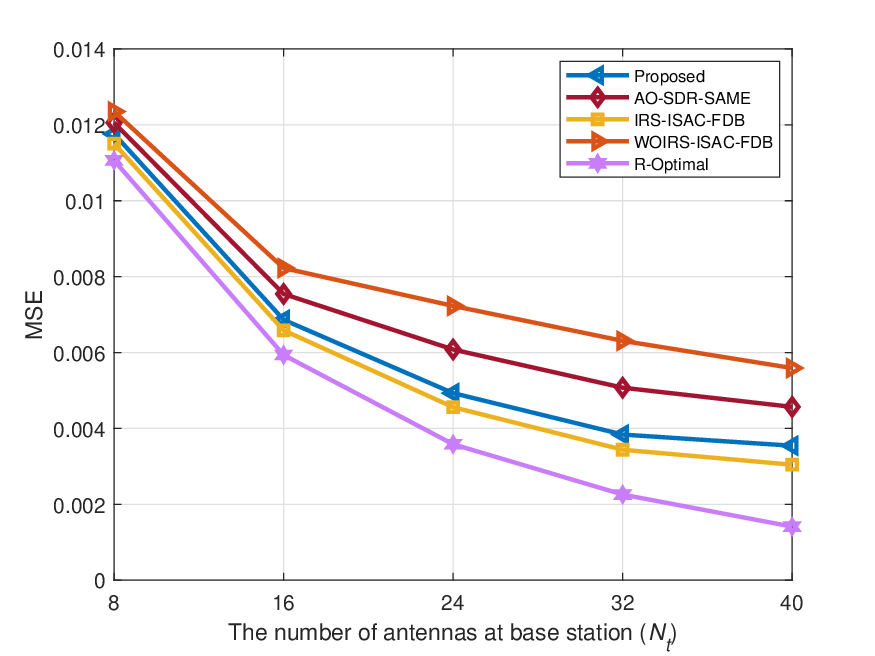} 
}
\subfigure[The MSE for the second case.]{
\includegraphics[width=3in]{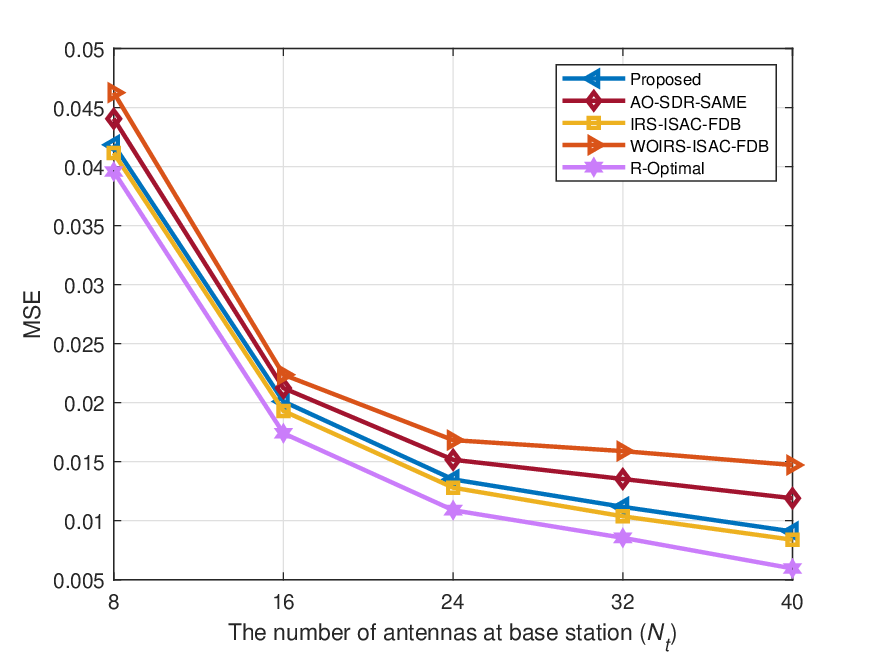} 
}
\caption{Comparison of MSE between different architectures and cases as the number of antennas at base station $(N_t)$ increases from 8 to 40.}\label{cMSE}
\end{figure}

To further evaluate radar performance, we increased the number of antennas at the base station $(N_t)$ and measured the MSE across different architectures, as shown in Figure \ref{cMSE}. As illustrated in Figures \ref{cMSE}(a) and \ref{cMSE}(b), the MSE performance of the proposed method is significantly lower than that of both the 'WOIRS-ISAC-FDB' architecture and the 'AO-SDR-SAME' architecture, and it is nearly identical to that of the 'IRS-ISAC-FDB' architecture. Specifically, in the first case, when $N_t=24$, the MSE of the proposed method is 0.0049, closely matching the MSE of the 'IRS-ISAC-FDB' architecture (0.0046) and considerably lower than that of the 'WOIRS-ISAC-FDB' architecture (0.0072) and the 'AO-SDR-SAME' architecture (0.0060). In the second case, when $N_t=24$, the MSE of the proposed method is 0.0135, nearly the same as that of the 'IRS-ISAC-FDB' architecture (0.0128) and much lower than that of the 'WOIRS-ISAC-FDB' architecture (0.0168) and the 'AO-SDR-SAME' architecture (0.0152). Additionally, as $N_t$ increases, the MSE gaps between the different architectures become larger, primarily because the system gains more degrees of freedom, enabling improved performance.

\subsection{Comparisons of Communication Performances}

\begin{figure}[t]
  \begin{center}
  \includegraphics[width=3in]{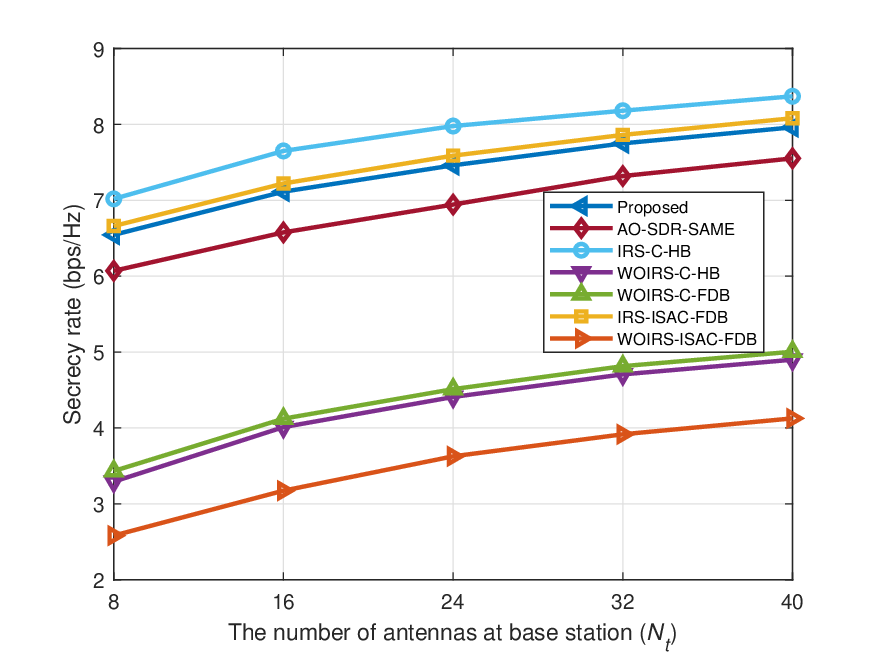}\\
  \caption{Comparison of secrecy rate between different architectures as the number of antennas at the base station $(N_t)$ increases from 8 to 40.}\label{cosrbd}
  \end{center}
\end{figure}

Figure \ref{cosrbd} illustrates the secrecy rate of different strategies as a function of the number of transmit antennas at the base station $(N_t)$. As shown in the figure, the secrecy rate improves with an increasing number of antennas. Additionally, the 'IRS-C-HB', 'IRS-ISAC-FDB', 'Proposed', and 'AO-SDR-SAME' architectures with IRS assistance achieve significantly higher secrecy rates than the 'WOIRS-C-HB', 'WOIRS-C-FDB', and 'WOIRS-ISAC-FDB' architectures. This substantial performance boost is attributed to the IRS, which provides greater degrees of freedom in system design. Notably, with IRS assistance, the proposed method can even outperform the communication-only architectures 'WOIRS-C-HB' and 'WOIRS-C-FDB'. Furthermore, the proposed architecture achieves nearly the same secrecy rate as the 'IRS-ISAC-FDB' architecture across varying values of $N_t$, while utilizing only one-fourth of the RF chains. Moreover, the proposed architecture, enhanced by the highly efficient distributed PDD algorithm, achieves a higher secrecy rate compared to the 'AO-SDR-SAME' architecture utilizing the AO method. Finally, the secrecy rate gap between the proposed architecture and the 'IRS-C-HB' architecture arises from the dual beamforming used for radar sensing.

\begin{figure}[t]
  \begin{center}
  \includegraphics[width=3in]{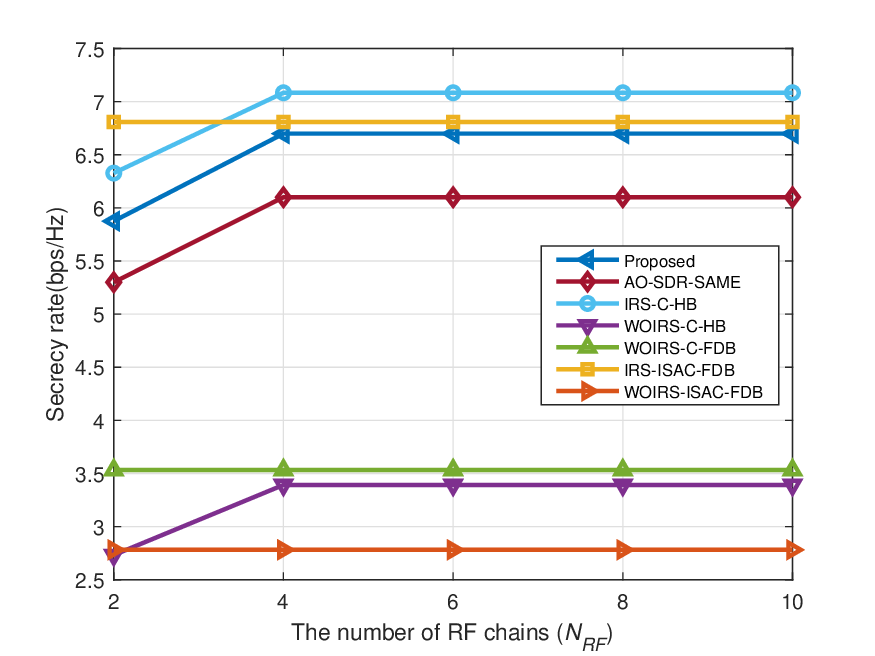}\\
  \caption{Comparison of secrecy rate between different architectures as the number of RF chains $(N_{RF})$ increases from 2 to 10.}\label{cosrbdas}
  \end{center}
\end{figure}

Figure \ref{cosrbdas} illustrates the secrecy rate of different strategies under varying numbers of RF chains $(N_{RF})$. It can be seen that the secrecy rate of the architectures with HB increases as the number of RF chains rises from 2 to 4. Beyond $N_{RF}>4$, the secrecy rate of these strategies stabilizes and becomes very close to those of the architectures with FDB. This result demonstrates that when $N_{RF}>2M$, the proposed architecture achieves a near-optimal solution, which is consistent with the findings in \cite{yu2016alternating}, highlighting the high efficiency of the proposed PDD-based algorithm. Additionally, it is interesting to note that the proposed architecture achieves a higher secrecy rate with $N_{RF}>4$ compared to the 'IRS-C-HB' architecture with $N_{RF}=2$. 

\begin{figure}[t]
  \begin{center}
  \includegraphics[width=3in]{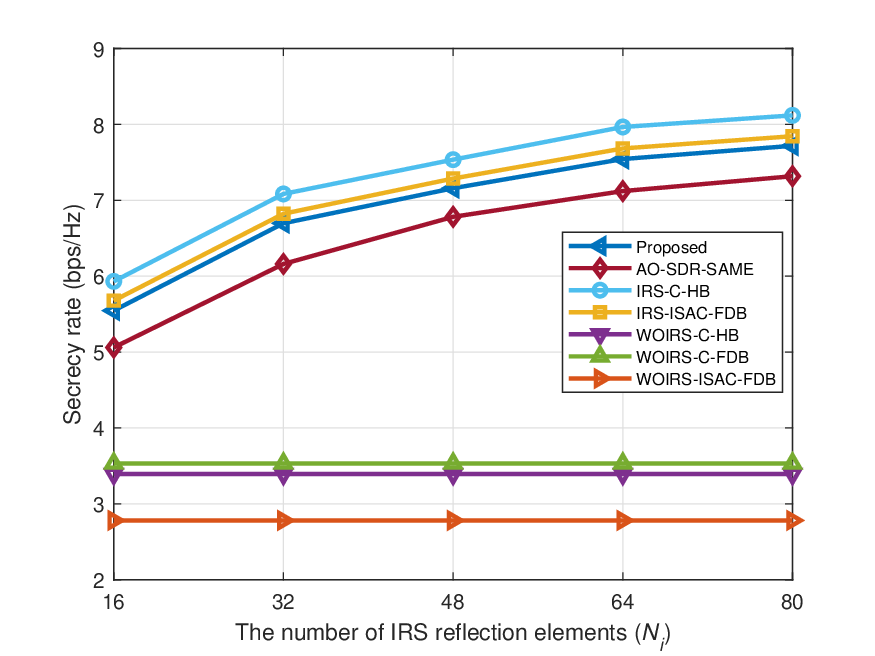}\\
  \caption{Comparison of secrecy rate between different architectures as the number of IRS reflection elements $(N_{i})$ increases from 16 to 80.}\label{corsire}
  \end{center}
\end{figure}

Figure \ref{corsire} illustrates the secrecy rate of different architectures as the number of IRS reflection elements $(N_{i})$ increases from 16 to 80. As seen in the figure, due to the additional degrees of freedom provided by the IRS, the IRS-assisted architectures ('Proposed', 'AO-SDR-SAME', 'IRS-C-HB', and 'IRS-C-FDB') achieve higher secrecy rates compared to the 'WOIRS-C-HB', 'WOIRS-C-FDB', and 'WOIRS-ISAC-FDB' architectures, even when the number of IRS reflection elements is as small as $N_{i}=16$. Additionally, the slope of the increase in secrecy rate for the IRS-assisted architectures ('Proposed', 'IRS-C-HB', 'IRS-C-HB', and 'IRS-C-FDB') slightly decreases, likely due to the increased difficulty in handling larger matrix sizes. Furthermore, the proposed architecture achieves a higher secrecy rate than the 'WOIRS-C-FDB' architecture, demonstrating that the IRS can greatly enhance the performance of the ISAC system.    

\begin{figure}[t]
  \begin{center}
  \includegraphics[width=3in]{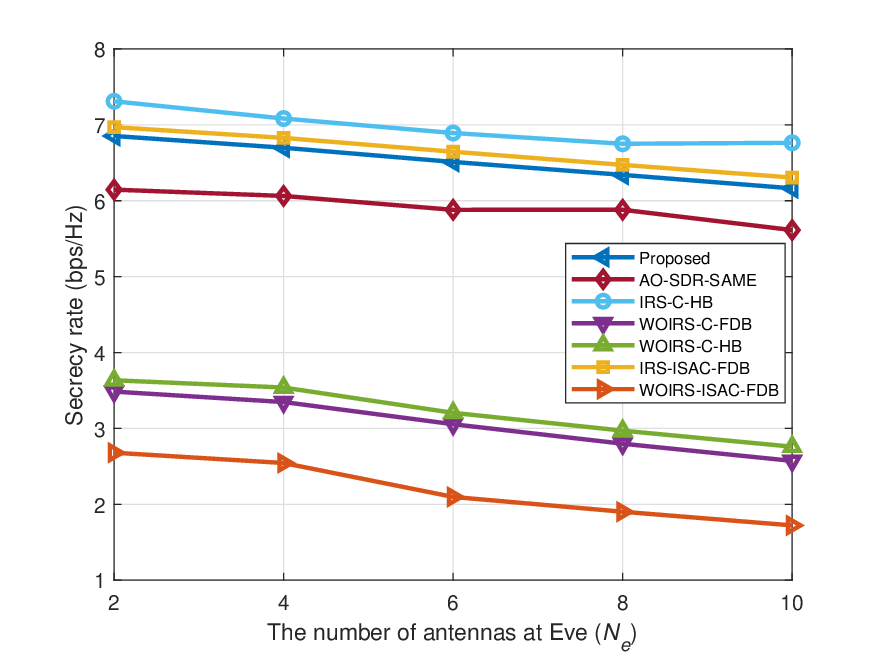}\\
  \caption{Comparison of secrecy rate between different architectures as the number of antennas at Eve $(N_{e})$ increases from 2 to 10.}\label{eiiii}
  \end{center}
\end{figure}

\begin{figure}[t]
  \begin{center}
  \includegraphics[width=3in]{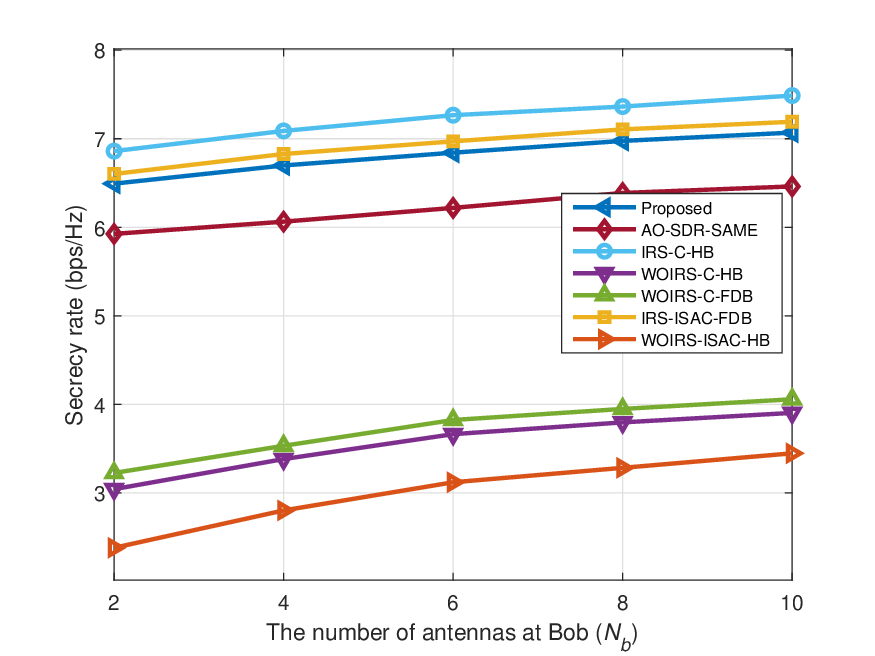}\\
  \caption{Comparison of secrecy rate between different architectures as the number of antennas at Bob $(N_{b})$ increases from 2 to 10.}\label{biiii}
  \end{center}
\end{figure}

Figures \ref{eiiii} and \ref{biiii} depict the secrecy rate of various architectures as the number of antennas at Eve $(N_{e})$ and Bob $(N_{b})$ increases from 2 to 10, respectively. These figures reveal contrasting trends. With an increase in $N_{e}$, the secrecy rate of these architectures declines. Conversely, as $N_{b}$ rises, the secrecy rate enhances. This is anticipated because more antennas enable both Eve and Bob to gather additional information. Specifically, in Figure 9, when $N_{e}=6$, the proposed method achieves a secrecy rate of 6.5114, which is almost identical to the 'IRS-ISAC-FDB' architecture at 6.6459. This rate is significantly higher than that of the 'AO-SDR-SAME' architecture at 5.8803, the 'WOIRS-C-FDB' architecture at 3.0556, the 'WOIRS-C-HB' architecture at 3.2055, and the 'WOIRS-ISAC-FDB' architecture at 2.0974. In Figure 10, when $N_{b}=6$, the proposed method attains a secrecy rate of 6.8428, nearly matching the 'IRS-ISAC-FDB' architecture at 6.9705. This performance is substantially better than the 'AO-SDR-SAME' architecture at 6.2193, the 'WOIRS-C-FDB' architecture at 3.8237, the 'WOIRS-C-HB' architecture at 3.6637, and the 'WOIRS-ISAC-FDB' architecture at 3.1203.

\section{Conclusion}
In this paper, we explore the use of hardware-efficient HB architecture in IRS-ISAC systems. Our focus is on enhancing PLS for communication and reducing beampattern similarity for radar sensing. The formulated problem is non-convex and challenging to solve. To address this efficiently, we propose an algorithm based on the PDD framework, featuring a closed-form solution at each step. Simulation results validate the effectiveness of the proposed algorithm and demonstrate the superiority of the IRS-ISAC system with HB architecture in balancing performance and hardware costs.

\begin{appendices} 
\section{Tight upper bound for (\ref{objevtivesubF})}
\label{appendixA}
We show that the function (\ref{maxsubF}) is a tight upper bound of the objective function (\ref{objevtivesubF}). To begin with, the objective function (\ref{objevtivesubF}) is equivalent to,
\begin{subequations}
\begin{align}
    \Re \{ \text{Tr}\{ {\bm \Psi}^H ( {\bf Q} & - {\bf F}{\bf W} ) \} \}  + \frac{1}{2\rho} \|  {\bf Q} - {\bf F}{\bf W}  \|_F^2 \\ &= \frac{1}{2\rho}|| {\bf F}{\bf W} - ( {\bf Q}+\rho{\bm \Psi} ) ||_F^2 \\ &= \Re \{ 
 \text{Tr}\{ {\bf F} {\bf W} {\bf W}^H{\bf F}^H - {\bf F}{\bf D} - {\bf F}^H{\bf D}^H  \}\} \\ &= \sum\limits_{i = 1}^{N_t} \Re \{ {\bf f}_i{\bf G}{\bf f}_i^H - 2{\bf f}_i{\bf d}_i \},\label{reformulateupperbunddetri}
\end{align}
\end{subequations}
where ${\bf f}_i \in \mathbb{C}^{1 \times N_{RF}}$ is the $i$-th row of the matrix $\bf F$; ${\bf G}={\bf W}{\bf W}^H \in \mathbb{C}^{N_{RF} \times N_{RF}}$; ${\bf D}={\bf W}({\bf Q} + \rho {\bf \Psi})^H= [{\bf d}_1,...,{\bf d}_i,...,{\bf d}_I] \in \mathbb{C}^{N_{RF} \times N_{t}}$.

We need the following lemma from \cite{song2015optimization} (\textit{we reuse the some notations in this lemma}),

\textit{Lemma 1 ([\textit{40}, Lemma 1])}: The quadratic function of the form ${\bf w}^H {\bf T} {\bf w}$, with ${\bf T}$ being a Hermitian matrix is majorized by ${\bf w}^H {\bf S} {\bf w}+2\text{Real}({\bf w}^H({\bf T}-{\bf S}){\bf w}^k)+({\bf w}^k)^H({\bf S}-{\bf T}){\bf w}^k$ at the point ${\bf w}^k$, where ${\bf S}$ is a Hermitian matrix such that ${\bf S}\succeq{\bf T}$.

\textit{Lemma 1} can be easily proven using second-order Taylor expansion and subsequently replacing the Hessian matrix ${\bf T}$ by another Hermitian matrix ${\bf S}$ such that ${\bf S}\succeq{\bf T}$. For a general twice differentiable function, \textit{Lemma 1} is also known by the name of the quadratic upper bound principle as mentioned in \cite{hunter2004tutorial}.

Following \textit{Lemma 1}, the tight upper bound for the first term of (\ref{reformulateupperbunddetri}) is given as,
\begin{equation}
\begin{aligned}
      {\bf f}_i{\bf G}{\bf f}_i^H  \le & \lambda_{\text{max}}({\bf G}) {\bf f}_i  {\bf f}_i^H \\
    +& 2\text{Real}\{{\bf f}_i({\bf G}-\lambda_{\text{max}}({\bf G}){\bf I})({\bf f}_i^k)^H\}\\
    +&{\bf f}_i^k(\lambda_{\text{max}}({\bf G}){\bf I}-{\bf G})({\bf f}_i^k)^H,   
\end{aligned}
\label{detriveupperboundfinal}
\end{equation}
where $\lambda_{\text{max}}({\bf G})$ denotes the maximum eigenvalue of $\bf G$. The first term on the right hand side of (\ref{detriveupperboundfinal}) is constant since due to the unit-modulus property of the variable $\bf F$, and the third term is also constant since it is independent of ${\bf f}_i$. After ignoring the constant term in (\ref{detriveupperboundfinal}), we are able to obtain a tight upper bound (\ref{maxsubF}). 

\section{Proof of Lemma \ref{lemma1}}
\label{appendixD}
The objective function (\ref{tightupperboundF }) can be rewritten as,
\begin{equation}
    f({\bf f}_i) = -2\Re \{ {\bf f}_i {\bf k} \}, \label{reforfk}
\end{equation}
where $ {\bf k} = \left(( \lambda_{\text{max}}({\bf G}) {\bf I}- {\bf G} ) ({\bf f}_i^k)^H + {\bf d}_i\right) $. Given that all entries of the vector ${\bf f}_i$ have constant modulus, the minimum of (\ref{reforfk}) is achieved when ${\bf f}_i$ is aligned with ${\bf k}$, leading to,
\begin{equation}
    {\bf f}_i^* = \text{exp}(j\text{arg} (  {\bf k} )  ). \label{optimumfi}
\end{equation}
We show that $ {\bf f}_i^* = \text{exp}(j\text{arg} (  {\bf k} )  )$ is the unique minimum, by contradiction. Let $ {\bf f}_i^+ = \text{exp}(j(\text{arg} (  {\bf k} ) + {\bm \iota}   ) )$ be another minimum, where at least one entry of ${\bm \iota}$ is non-zero and is not an integer multiple of $2\pi$ (otherwise, ${\bf f}_i^*={\bf f}_i^+$). Then, from (\ref{optimumfi}), we have,
\begin{subequations}
    \begin{align}
  &  f({\bf f}_i^*) = -2 \sum_{j=1}^{N_{RF}} | {\bf k}|_j, \label{subfi1}\\
  &  f({\bf f}_i^+) = -2 \sum_{j=1}^{N_{RF}} | {\bf k}|_j\cos({\bm \iota}_j).\label{subfi2}
\end{align}
\end{subequations}
Since both ${\bf f}_i^*$ and ${\bf f}_i^+$ minimize ${\bf f}_i^*$, it follows from (\ref{subfi1}) and (\ref{subfi2}) that, $f({\bf f}_i^*)=f({\bf f}_i^+)$. This is possible only if $\cos({\bm \iota}_j) = 1$, which implies that all the entries of ${\bm \iota}$ are zero or some integer multiples of $2\pi$ going against the assumption. This leads to a contradiction and hence proving the uniqueness. Based on the above discussion, it is clear to show that (\ref{solveF}) is the unique optimal solution of the subproblem (\ref{subreforF}).

\section{Tight upper bound for (\ref{objecsubQ})}
\label{appendixB}
To begin with, the second term of (\ref{objecsubQ}) is equivalent to,
\begin{subequations}
    \begin{align}
      \frac{(1-\mu)}{K}\sum\limits_{k = 1}^{K} & {{| {\delta P_d(\theta_k) - {\bf a}^H( {\theta}_k ){\bf Q} {\bf Q}^H{\bf a}( {\theta}_k )} |}^2} \\ = \frac{(1-\mu)}{K}\sum\limits_{k = 1}^{K}& \Re\{ \text{vec}^H({\bf Q}{\bf Q}^H) \text{vec}({\bf A}_k)\text{vec}^H({\bf A}_k)\text{vec}({\bf Q}{\bf Q}^H) \notag \\ &- 2\delta P_d(\theta_k)\text{vec}^H({\bf Q}{\bf Q}^H)\text{vec}({\bf A}_k) \} \\ =\Re\{ \text{vec}^H({\bf Q}&{\bf Q}^H){\bf C}\text{vec}({\bf Q}{\bf Q}^H) - \text{vec}^H({\bf Q}{\bf Q}^H){\bf b}_t \}, \label{finalequavilentQ}
\end{align}
\end{subequations}
where,
\begin{subequations}
    \begin{align}
        &{\bf A}_k = {\bf a}( {\theta}_k ){\bf a}^H( {\theta}_k ) \in \mathbb{C}^{N_t^2 \times N_t^2},\\ 
        &{\bf C} = \frac{(1-\mu)}{K}\sum\limits_{k = 1}^{K}\text{vec}({\bf A}_k)\text{vec}({\bf A}_k)^H \in \mathbb{C}^{N_t^2 \times N_t^2},\label{subupperQC}\\&{\bf b}_t = \frac{(1-\mu)}{K}\sum\limits_{k = 1}^{K}2\delta P_d(\theta_k)\text{vec}({\bf A}_k) \in \mathbb{C}^{N_t^2 \times N_t^2}.
    \end{align}
\end{subequations}
Similar to APPENDIX \ref{appendixA}, following \textit{Lemma 1}, a tight upper bound for the first term of (\ref{finalequavilentQ}) is given as,
\begin{subequations}
\begin{align}
 \text{vec}^H({\bf Q}&{\bf Q}^H){\bf C}\text{vec}({\bf Q}{\bf Q}^H) \\ 
\le & \lambda_{\text{max}}({\bf C})\text{vec}^H({\bf Q}{\bf Q}^H)\text{vec}({\bf Q}{\bf Q}^H) \notag\\ &+ 2 \Re \{ \text{vec}^H({\bf Q}{\bf Q}^H)({\bf C}-\lambda_{\text{max}}({\bf C}){\bf I})\text{vec}({\bf Q}^k({\bf Q}^k)^H) \} \notag  \\&+ \text{vec}^H({\bf Q}^k({\bf Q}^k)^H) ( \lambda_{\text{max}}({\bf C}){\bf I} -{\bf C})\text{vec}({\bf Q}^k({\bf Q}^k)^H)
\label{detrivemiddleQC1} \\ \le & \Re \{ \text{vec}^H({\bf Q}{\bf Q}^H) ({\bf c}_1-{\bf c}_2) \} \label{detrivemiddleQC2}\\
= & \Re \{ \text{Tr}\{ {\bf Q}^H {\bf C}_1 {\bf Q} \} \} - \Re \{ \text{Tr}\{ {\bf Q}^H {\bf C}_2 {\bf Q} \} \},
\end{align}
\label{useuppervecc}
\end{subequations}
where,
\begin{subequations}
\begin{align}
{\bf c}_1 &= \text{vec}({\bf C}_1) = 2{\bf C}\text{vec}({\bf Q}^k({\bf Q}^k)^H) \in \mathbb{C}^{N_t^2},\\
{\bf c}_2 &= \text{vec}({\bf C}_2) = 2\lambda_{\text{max}}({\bf C})\text{vec}({\bf Q}^k({\bf Q}^k)^H) \in \mathbb{C}^{N_t^2}.
\end{align}
\label{useupperc}
\end{subequations}
Notice that in (\ref{detrivemiddleQC1}), the first term is constant because $||{\bf Q}||_F^2 \le P_{max}$, and the third term is also constant since it is independent of $\bf Q$. Then after ignoring the constant term, (\ref{detrivemiddleQC1}) is converted into (\ref{detrivemiddleQC2}). Then based on (\ref{useuppervecc}) and (\ref{useupperc}), the first and second terms of objective function (\ref{objecsubQ}) can be reformulated as,
\begin{subequations}
    \begin{align}
       {\cal L}_{\rho} ({\bf Q}) \le & \mu(||{\bf H}_e({\bf \Phi }) {\bf Q}||_F^2-\eta||{\bf H}_b({\bf \Phi }) {\bf Q}||_F^2) \notag \\ &+ \Re \{ \text{Tr}\{ {\bf Q}^H {\bf C}_1 {\bf Q} \} \} - \Re \{ \text{Tr}\{ {\bf Q}^H ({\bf C}_2 + {\bf B}_t ){\bf Q} \} \} \\
       = & \Re \{ {\bf Q}^H {\bf Z}_1 {\bf Q} \} + \Re \{ {\bf Q}^H {\bf Z}_2 {\bf Q} \}, \label{twiceQuppermax}
    \end{align}
    \label{onceupperQ}
\end{subequations}
where ${\cal L}_{\rho} ({\bf Q})$ represents the first and second terms of objective function (\ref{objecsubQ}); ${\bf Z}_1 = \mu {\bf H}_e^H({\bf \Phi }){\bf H}_e({\bf \Phi })+{\bf C}_1 \in \mathbb{C}^{N_t \times N_t}$; ${\bf Z}_2 = -\mu\eta{\bf H}_b^H({\bf \Phi }){\bf H}_b({\bf \Phi })-{\bf C}_2 - {\bf B}_t \in \mathbb{C}^{N_t \times N_t}$; ${\bf b}_t = \text{vec}({\bf B}_t)$. Notice that (\ref{twiceQuppermax}) is still non-convex due to the matrix ${\bf Z}_2$ is negative semi-definite. By exploiting the ﬁrst-order Taylor expansion, a convex tight upper
bound of $\Re \{ {\bf Q}^H {\bf Z}_2 {\bf Q}\}$ is drived as, 
\begin{subequations}
    \begin{align}
        \Re \{ {\bf Q}^H {\bf Z}_2 {\bf Q}\} &= \sum\limits_{i = 1}^{M} {\bf q}_i^H {\bf Z}_2 {\bf q}_i\\ & \le \sum\limits_{i = 1}^{M} ( ({\bf q}_i^k)^H {\bf Z}_2 {\bf q}_i^k + 2 \Re\{ ({\bf q}_i^k)^H {\bf Z}_2 ({\bf q}_i - {\bf q}_i^k )\} ) \\ & \le \sum\limits_{i = 1}^{M} 2 \Re\{ ({\bf q}_i^k)^H {\bf Z}_2 {\bf q}_i \}\\&= \Re \{ \text{Tr} \{ {\bf Q}^H {\bf Z}_3 \} \},
    \end{align}
    \label{twiceupperQ}
\end{subequations}
where ${\bf Z}_3 = 2 {\bf Z}_2^H {\bf Q}^k \in \mathbb{C}^{N_t \times M}$. Finally, based on (\ref{onceupperQ}) and (\ref{twiceupperQ}), the majorized objective function (\ref{objectsubQ}) is therefore established. 

\section{Tight upper bound for (\ref{maxPhiobj})}
\label{appendixC}
The first term of (\ref{maxPhiobj}) is a quadratic function suitable fit in \textit{Lemma 1}, where its tight upper bound can be found same as APPENDIX \ref{appendixA}. Here, we omit the detailed derivation.

\end{appendices}

\bibliographystyle{IEEEtran}
\bibliography{Bibliography}

\vfill

\end{document}